\documentclass[
reprint,
 amsmath,amssymb,
 aps,
prb,
]{revtex4-2}

\usepackage{graphicx}
\usepackage{dcolumn}
\usepackage{bm}
\usepackage[T1]{fontenc}
\usepackage{xcolor}
\usepackage{siunitx}

\begin{document}

\title{Phase diagram of one-dimensional driven-dissipative  exciton-polariton condensates}

\author{Francesco Vercesi$^1$, Quentin Fontaine$^2$, Sylvain Ravets$^2$, Jacqueline Bloch$^2$, Maxime Richard$^{3,4}$, L\'eonie Canet$^{1,5}$, Anna Minguzzi$^1$}
\affiliation{$^1$ Université Grenoble Alpes, CNRS, LPMMC, 38000 Grenoble, France, $^2$ Universit\'e Paris-Saclay, CNRS, Centre de Nanosciences et de Nanotechnologies (C2N), 91120 Palaiseau, France, $^3$ Majulab International Research Laboratory, French National Centre for Scientifique Research, National University of Singapore, Nanyang Technological University, Sorbonne Universit\'e, Universit\'e C\^ote d'Azur, 117543 Singapore, $^4$ Centre for Quantum technologies, National University of Singapore, 117543 Singapore, $^5$ Institut Universitaire de France, 5 rue Descartes, 75005 Paris}

\begin{abstract}
We consider a one-dimensional driven-dissipative exciton-polariton condensate under incoherent pump, described by  the stochastic generalized Gross-Pitaevskii equation. It was shown that the condensate phase dynamics maps under some assumptions to the Kardar-Parisi-Zhang (KPZ) equation, 
and the temporal coherence of the condensate follows a stretched exponential decay characterized by KPZ universal exponents. In this work, we determine the main mechanisms which lead to the departure from the KPZ phase, and identify three possible other regimes: (i) a soliton-patterned regime at large interactions and weak noise,  populated by localized structures analogue to dark solitons; (ii) a vortex-disordered regime at high noise and weak interactions,  dominated by point-like phase defects in space-time; (iii) a defect-free reservoir-textured regime where the adiabatic approximation breaks down. We characterize each regime by the space-time maps, the first-order correlations, the momentum distribution and the density of topological  defects. We thus obtain the phase diagram at varying noise, pump intensity and interaction strength. Our predictions are amenable to observation in state-of-art experiments with exciton-polaritons.

\end{abstract}


\maketitle


\section{Introduction}
\label{sec:introduction}

Exciton-polaritons (polaritons) are photons-like quasi-particles that can behave, as a  many-body system, as a quantum fluid made of light \cite{Carusotto2013}. They are obtained in semiconductor optical microcavities when photons in the cavity mode are in the strong coupling regime with electron-hole excitations within the cavity medium, known as excitons \cite{Weisbuch1992}. Owing to the cavity short lifetime, an external excitation is required to replenish the system and achieve a non-equilibrium  steady-state. Under incoherent excitation, a striking phenomenon exhibited by these systems, is the nonequilibrium analogue of Bose-Einstein condensation first reported in Ref.~\cite{Kasprzak2006}. Several other paradigmatic features of quantum fluids have been reported in polariton quantum fluids such as superfluidity \cite{Amo2009}, solitonic excitations \cite{Amo2011, Barland2012} and quantized vortices \cite{Lagoudakis2008}.

In the non-equilibrium condensation mechanism, upon crossing the excitation intensity threshold, the macroscopic phase of exciton-polariton condensates is spontaneously chosen and owing to the incoherent nature of the excitation, remains free to fluctuate \cite{Carusotto2013}. A major difference of this condensate with respect to its equilibrium counterpart, is that it can display a critical behavior described by the Kardar-Parisi-Zhang equation \cite{Kardar1986}, a non-linear Langevin equation which captures the universal dynamics of an extremely large collection of non-equilibrium systems, including stochastically growing interfaces, randomly stirred fluids, directed polymers in random media and many more \cite{Takeuchi2018}. Indeed, it was established  in recent experiments \cite{Fontaine2022} that in one dimension and for condensates of  negative effective mass, the condensate phase exhibits the universal scaling behavior of the KPZ universality class, as was predicted in \cite{He2015, Ji2015}. These fluctuations determine the coherence in space and time of the exciton-polariton condensate, characterized by stretched exponential decays. The emergence of KPZ universal scaling was demonstrated experimentally thanks to precise interferometric measurements of spatial and temporal coherence, and supported by numerical modeling of the condensate evolution \cite{Fontaine2022}. Besides, numerical simulations of the microscopic models showed that the phase fluctuations probability distributions - non-Gaussian and skewed - match the characteristic distributions of KPZ stochastic processes \cite{Squizzato2018, Deligiannis2021}. 

The effective description of polaritons driven-dissipative condensates is given by the stochastic generalized Gross-Pitaevskii equation (gGPE) \cite{Carusotto2013}, a semiclassical nonlinear model in which drive and loss enter as complex coefficients and a white noise accounts for stochastic fluctuations intrinsically due to drive and dissipation. Under some conditions detailed in Ref.~\cite{Bobrovska2014}, its deterministic part is equivalent to the complex Ginzburg-Landau equation (CGLE), which has been widely studied both analytically and numerically \cite{Aranson2002}. Its rich phase diagram has been shown to exhibit nonlinear localization, solitons, pattern selection and spatio-temporal chaos \cite{Aranson2002, Chate1994, Chat1995}. Moreover, in Ref.~\cite{Grinstein1996} the large scale effective dynamics of the CGLE in the regime of phase turbulence (i.e. in absence of defects) was predicted to be controlled by the KPZ equation.

A rich variety of dynamical regimes can also be accessed in polaritons quantum fluids, by tuning the parameters of the gGPE. For instance, soliton generation and propagation has been predicted and experimentally observed both in one- and in two-dimensional geometries \cite{Amo2011, Carusotto2013, Pernet2022},  dynamical instabilities and defects nucleation preventing extended coherence have been studied in Refs.~\cite{Bobrovska2014, Bobrovska2015, Bobrovska2017, Baboux2018, Bobrovska2019}, while the dynamics of dark solitons was characterized in Refs.~\cite{Smirnov2014, Opala2018}.
In Ref.~\cite{He2017}, by solving a compact KPZ equation as well as the stochastic CGLE, it was shown that the KPZ coherence of one-dimensional driven-dissipative condensates can be disrupted due to the compact nature of the condensate phase, which can host spatio-temporal vortices. The phase diagram at varying noise strength and KPZ nonlinearity was investigated, and a vortex-turbulent regime at large effective KPZ nonlinearity was identified by analyzing the momentum distribution decay.

In this work, we determine the full phase diagram of a one-dimensional incoherently pumped polariton condensate, in a parameter space constituted by three key experimentally relevant parameters: the pump intensity, the blueshift of the condensate energy - related to the exciton-exciton interaction strength, and the noise amplitude. We first identify the extent of the KPZ regime for conditions close to the experiments of Ref.~\cite{Fontaine2022}, retrieving the KPZ scaling in the decay of the coherence. We then then identify the main mechanisms of departure from this regime, and report the emergence of three other regimes. The first one sets in the  low-noise regime at increasing blueshift  and is characterized by the stochastic proliferation and dynamics of solitonic structures. It is sustained by the dynamical instability that favor the development of long-lived high-contrast defects in the condensate. The second one emerges in the high-noise and moderate pump regime, and is characterized by the stochastic formation of defects that form quantized phase vortices in the space-time plane. 
In this regime, the phase-phase correlation function grows linearly in time and  deviates from the first-order correlation function. The latter 
still displays the KPZ scaling. In fact, within one defect and the next one, the phase of the condensate fluctuates according to KPZ scaling, but the presence of vortices add large jumps to the phase trajectories. Hence, this vortex regime regime may be viewed as piece-wise KPZ.
The third one appears in the high-pump and moderate-noise regime, and corresponds to the breakdown of the adiabatic approximation, and of the mapping to the KPZ equation. 
In this regime, the dynamics of the reservoir cannot be neglected and leads to larger fluctuations of both the reservoir and the condensate densities. In this regime, the KPZ scaling is washed out, even though the phase remains defect-free. 

For each three regimes, as well as for the KPZ one,  we show a typical time evolution, describe its salient features and calculate the first-order correlation function within the steady-state to determine the decay of the coherence. We also determine the momentum distribution, as well as the phase-phase correlations to characterize the effective phase dynamics. All these studies are eventually merged into a single phase diagram of a one-dimensional polariton condensate, within a parameter space directly relevant to experiments.

\section{Model}
\label{sec:model}

We consider a one-dimensional exciton-polariton condensate coupled to an excitonic reservoir which is pumped by an external optical excitation, and that feeds particles into the condensate via spontaneous and stimulated electronic scattering. The dynamics of the condensate wavefunction $\psi(x,t)$ and of the exciton reservoir density $n_R(x,t)$ are described, respectively, by the generalized stochastic Gross-Pitaevskii equation and by the rate equation  \cite{Carusotto2013}:
\begin{align}
     i\hbar \partial_t &\psi(x, t)= 
     \Big[\epsilon(\hat{k}) + g|\psi(x, t)|^2 + 2g_Rn_R(x,t)  \nonumber\\
     &+  \frac{i \hbar}{2}\left( R n_R(x,t) - \gamma(\hat{k}) \right) \Big]\psi(x,t) + \hbar \eta(x, t)
\label{eq:sGPE}\\
 \partial_t &n_R(x, t) = P(x)- \left(\gamma_R + R|\psi(x,t)|^2\right)n_R(x, t)
    \label{eq:reservoir}
\end{align}
 with $\hat{k}=-i\partial_x$. The pump $P(x)$ describes the laser excitation of the reservoir, which we consider to be homogeneous $P(x)=P$, $\gamma_R$ is the rate for spontaneous decay of excitons and $R n_R>0$ the exciton scattering rate into the condensate,  representing the condensate drive under incoherent pumping. 
Losses are taken into account through the $k$-dependent linewidth $\gamma(\hat{k})$. Its momentum dependence around $k=0$ is accounted for in a parabolic approximation as $\gamma(k) = \gamma_0 + \gamma_2 k^2$, where $\gamma_0^{-1}$ represents the polariton lifetime and $\gamma_2>0$ is a diffusion constant.

It was shown in Ref.~\cite{Baboux2018} that a positive effective mass leads to fragmentation due to self-focusing instabilities, hampering large spatial coherence.
Hence, 
 like in the experiments of Ref.~\cite{Fontaine2022}, we add a periodic potential in order to achieve an effective negative mass for polaritons. This potential results in a polaritonic dispersion relation that we model effectively via the kinetic term $\epsilon(k) = E_0 + 2J\mathrm{cos}(k a)$, where $a$ is the lattice spacing, and in which a condensate forms at $k=0$. The polaritons effective mass around $k=0$ is thus negative, ensuring effective repulsive polariton-polariton interactions in presence of the reservoir.
 In the experiment, polaritons are in a Lieb lattice providing them with an effective mass around the condensate state of $m_{exp} = -3.3 \cdot 10^{-6} m_e$ \cite{Fontaine2022}, where $m_e$ is the electron mass. We thus set $J = \hbar^2/(2|m_{exp}|a^2)$ and choose $E_0 = -2J$ such that the reference energy is zero at $k=0$.

The condensate gain and loss processes result in stochastic fluctuations which are described by a complex Gaussian noise $\eta(x,t)$, with 
\begin{align}
    & \langle \eta(x, t) \rangle = 0 \\
    & \langle \eta(x, t) \eta(x', t')\rangle = 2 \sigma\delta(x-x')\delta(t-t').
\label{eq:noise}
\end{align}
In the microscopic model, the fluctuations amplitudes are determined as $\sigma_0 = \frac{1}{2}\left(Rn_R + \gamma \right)$ \cite{Wouters2009}. In this work we assume that additional fluctuation sources may be present, or added in a controlled way to the system by suitable noise engineering. We henceforth consider the noise amplitude $\sigma$ as an adjustable parameter, and also allow for values smaller that the nominal noise $\sigma_0$ to obtain a complete view of the phase diagram.

It is useful to recall the homogeneous mean-field solution of equations (\ref{eq:sGPE})  and (\ref{eq:reservoir}), given by $n_{R0} = \frac{\gamma_0}{R}$ and $\psi_0(t) = \sqrt{\rho_0} \mathrm{e^{-i\mu_0 t}}$, with $\rho_0 = \frac{\gamma_R}{R} (p-1)$.
Here
$p = \frac{P}{P_{th}} = \frac{P R}{\gamma_0 \gamma_R}$
is the reduced pump, where $P_{th}$ is the threshold pump intensity for condensation.
The blueshift due to the two-body interactions is given by $\mu_0 = g\rho_0 + 2g_Rn_{R0}$, with $g,g_R\geq 0$.
Since we aim to stay close to the conditions of the experiments of Ref.~\cite{Fontaine2022}, in which the interactions are dominated by the reservoir of excitons, we take $g=0$ all through the paper.
In this case $\mu \simeq \mu_{\rm th} = 2g_Rn_{R0} $, the blueshift of the polariton spectrum when the pump is tuned at threshold value (i.e. $\rho=0$). 

Under certain assumption discussed below, starting from the stochastic Gross-Pitaevskii equation (\ref{eq:sGPE}), the dynamics of the phase $\theta$ can be mapped \cite{He2015, Ji2015}  to  the  one-dimensional Kardar-Parisi-Zhang equation \cite{Kardar1986}:
\begin{equation}
    \partial_t \theta = \nu \partial^2_x \theta + \frac{\lambda}{2}\left( \partial_x \theta\right)^2 + \sqrt{D}\xi .
    \label{eq:kpz_phase}
\end{equation}
The KPZ parameters are directly related to the microscopic ones according to  $\nu = \frac{1}{2} \left(\alpha\frac{\hbar}{2m} + \gamma_2 \right), \lambda = -2\left( \frac{\hbar}{2m} - \alpha \gamma_2 \right), D = \frac{\sigma}{2\rho_0}(1+4a^2)$ and $\alpha = \frac{(\gamma_R + R\rho_0)^2}{R^2 P}\left(g - \frac{2g_RPR}{(\gamma_R + R\rho_0)^2} \right)$ \cite{Fontaine2022}.  Equation (\ref{eq:kpz_phase}) holds assuming that the density fluctuations of both the condensate and the reservoir and their spatial variations are negligible as compared to the phase ones. Also, the mapping is valid as long as the phase is a slowly varying function in space and time. We show in the following that these assumptions break down in various cases, leading to the formation of a soliton-patterned,  a disordered-vortex, or a reservoir-textured regime.

\section{Characterization : observables and methods}

\subsection{First-order coherence, phase correlations and momentum distribution}
We characterize the coherence of the condensate by computing the first-order space-time correlation function
\begin{equation}
    g^{(1)}(\Delta x, \Delta  t) = \frac{\langle \psi^*(x_0+\Delta x, t_0+\Delta t) \psi(x_0, t_0) \rangle} {\sqrt{\langle |\psi(x_0+\Delta x, t_0+\Delta t)|^2 \rangle \langle |\psi(x_0, t_0)|^2 \rangle}},
\end{equation}
where $\langle..\rangle$ denotes the average over the stochastic trajectories generated by the noise $\eta$.
 Upon assuming that density-density and density-phase correlations are negligible, the behavior of the coherence is determined by the phase-phase correlations, as follows from the  cumulant expansion of  $g^{(1)}(\Delta x,\Delta t)$, which gives to lowest order
\begin{equation}
    |g^{(1)}(\Delta x,\Delta t)| \approx \mathrm{e}^{-\frac{1}{2}C_{\theta}(\Delta x, \Delta t)}
    \label{eq:g1_approx_Ctheta}
\end{equation}
with $C_{\theta}(\Delta x,\Delta t) = \langle \left[\theta(x_0+\Delta x, t_0+\Delta t) - \theta(x_0, t_0) \right]^2\rangle$.

In the KPZ regime the phase correlation function $C_{\theta}(\Delta x,\Delta t)$  is predicted to display the universal scaling form \cite{Kardar1986}:
\begin{align}
    C_{\theta}^{(\rm KPZ)}(\Delta x,\Delta t)& = C_0 \Delta t^{2\beta} g^{(\rm KPZ)}\left(\alpha_0 \frac{\Delta x}{\Delta t^{1/z}}\right) \nonumber\\
     & \sim \begin{cases}
    |\Delta x|^{2\chi} &t\rightarrow0 \\
   \Delta t^{2\beta} &x\rightarrow0
    \end{cases}
    \label{eq:_Ckpz_scalingform}
\end{align}
where $\chi=1/2$, $\beta=1/3$, $z=\chi / \beta = 3/2$ are the exact universal Kardar-Parisi-Zhang exponents in one dimension, $g^{(\rm KPZ)}(y)$ is the universal scaling function calculated exactly in Ref.~\cite{Prhofer2004}, and $C_0, \alpha_0$ are non-universal normalization parameters.

Equal-time correlations are also used to obtain the polariton momentum distribution, given by   

\begin{equation}
n(k)=\langle \psi^*(k, t_0) \psi(k, t_0) \rangle 
\label{eq:nk}
\end{equation}
where $\psi(k, t )= \int dx \, e^{ik x}\psi(x, t)$.

\subsection{Soliton and vortex counting}
When varying the parameters of the gGPE, we evidence various types of topological defects, in particular dark or grey solitons and space-time vortices.
Recalling that a dark soliton is characterized by a vanishing density at its core and a $\pi$ jump in the phase \cite{Leggett2004}, 
we estimate the soliton density by 
\begin{equation}
    n_s(t) = \frac{1}{\pi L} \int_0^L |\partial_x \theta(x, t)| dx,
    \label{eq:nu}
\end{equation}
For a soliton train made of $N_s$ dark solitons with cores at positions $x_n$ one would have $|\partial_x \theta(x)|\simeq \pi \sum_{n=1}^{N_s} \delta(x-x_n)$ 
 hence $ L n_s= \int_0^L \sum_{n=1}^{N_s} \delta(x-x_n) dx = N_s$. For the case of grey solitons, the above estimator yields $L n_s < N_s$, still providing a reliable estimate of the soliton density,  especially at low noise. Note that this estimator also counts the  phase variations due to the noise perturbation, contributing a background value which becomes important at large noise.  In the numerical calculations, we evaluate the phase  derivative as
    $\partial_x \theta(x, t) = -i  e^{-i\theta(x,t)} \partial_x \, e^{i\theta(x,t)}$ for better accuracy.

As discussed e.g. in Ref.~\cite{He2017},  vortex defects in space time  may appear in 1D quantum systems.
We account for space-time vortices by considering the $z$-component of the space-time vorticity field $\Omega(x,t) = \partial_x \partial_t \theta(x,t) - \partial_t \partial_x \theta(x, t)$.
The flux of $\Omega(x,t)$ through the surface $L \times T$ gives a multiple of $2\pi$, allowing to estimate the total density of vortices as:
\begin{equation}
    n_v = \frac{1}{2\pi LT} \int_0^T \int_0^L |\Omega(x,t)|dtdx
    \label{eq:number-vortices}
\end{equation}
which accounts for the positively and negatively charged vortices.
Note that, although the two types of topological defects are markedly different, the two indicators defined in \eqref{eq:nu} and \eqref{eq:number-vortices} share some similarities  and provide comparable estimates for the number of defects. 

\subsection{Bogoliubov stability diagram}
\begin{figure}
    \includegraphics[width=.4\textwidth]{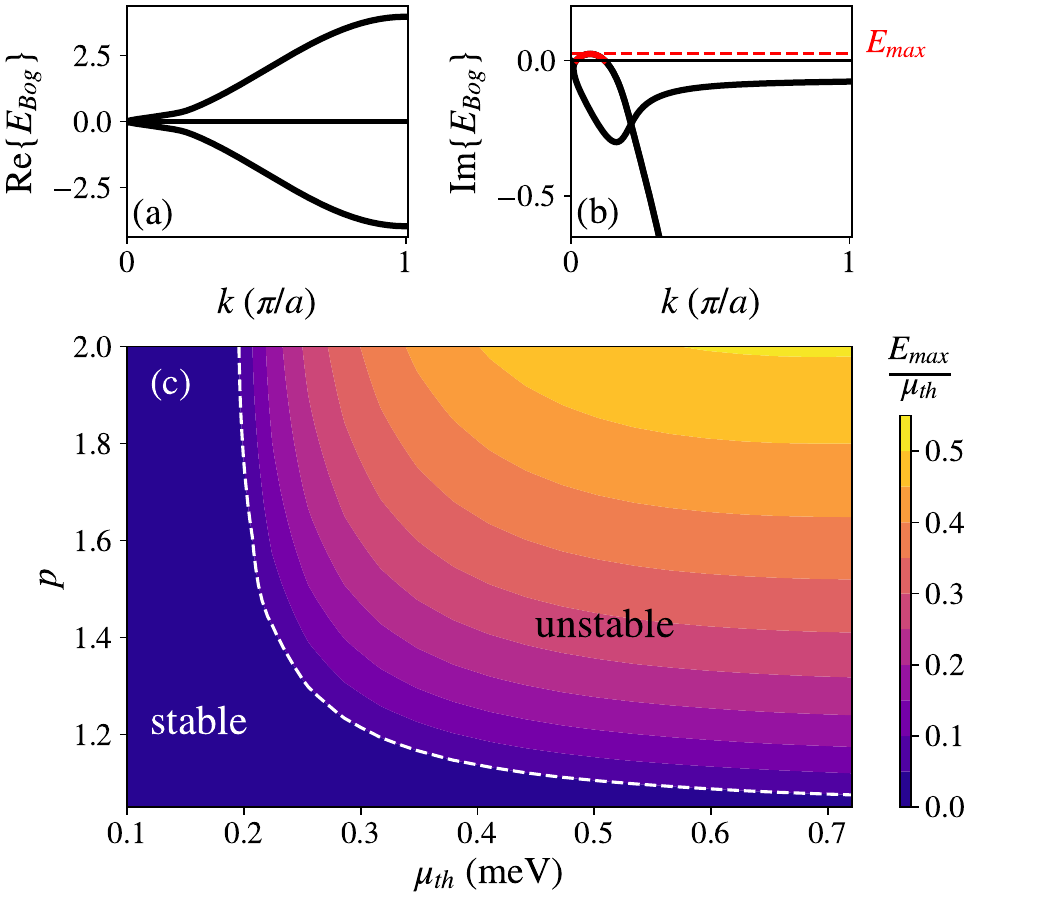}
    \caption{(Color online)
   (a, b) Real and imaginary parts of the Bogoliubov spectrum for $p=1.15$, $\mu_{\rm th}=0.6$ \si{meV}. (c) Bogoliubov stability diagram of Eqs. (\ref{eq:sGPE}, \ref{eq:reservoir}) in the plane  $(\mu_{\rm th},p)$. The instability is given by the condition $E_{max} = \textrm{max}\{\textrm{Im}(E_{\rm Bog}(k))\}>0$.}
    \label{fig:bogo_phasediagram}
\end{figure}

The development of  contrasted structures in the condensate, out a weak incoming noise, suggests that the condensate is in a regime in which its fluctuations are unstable. We thus determine the Bogoliubov spectrum of the condensate elementary excitations $E_{\rm Bog}(k)$, which describes the linearized coupled dynamics of small fluctuations of condensate and reservoir density around the homogeneous mean-field solution \cite{Carusotto2013}.  For each given set of parameters, we detect the presence of modulational instabilities when the imaginary part of the Bogoliubov spectrum is positive \cite{Carusotto2013} (see Fig.~\ref{fig:bogo_phasediagram}). We follow the maximum value of Im $\left\{ E_{\rm Bog}(k)\right\}$ to track the unstable regions in parameter space. The resulting Bogoliubov stability diagram in the $(\mu_{\rm th}, p)$ plane is illustrated  in Figure \ref{fig:bogo_phasediagram}.

We remark that the origin of the instability is not due to a change in sign of the effective two-body coupling strength $g_{\rm eff} = g - \frac{2g_R\gamma_0}{\gamma_R p}$ since as $g=0$, it is always negative in our case. If it were the case, the stable and unstable areas in Fig.~\ref{fig:bogo_phasediagram} would be separated by the straight line defined by $g_{\rm eff}=0$.  The instability shown here is rather connected with the breakdown of adiabaticity of the reservoir dynamics (\ref{eq:reservoir}), similarly to \cite{Bobrovska2015}.

For later comparison, we also calculate the momentum distribution within the linear Bogoliubov approximation, in the limit of adiabatic evolution of the reservoir. 
\begin{equation}
n_k = 2 \pi |\psi_0|^2 \delta(k) + \frac{\sigma}{\Gamma_0 + \Gamma_k}\left(1 + \frac{\Gamma_0^2 + \tilde{\mu}^2}{E_k^2 + \Gamma_k^2 + 2\Gamma_0\Gamma_k} \right)
\end{equation}
where  $\Gamma_0 = \frac{1}{2}\gamma_0 \left(\frac{p-1}{p}\right)$, $\Gamma_k = \frac{1}{2} \gamma_2k^2$, $E_k^2 = \epsilon_k(\epsilon_k + 2\tilde{\mu})$ and $\tilde{\mu} = g_{\rm eff}|\psi_0|^2$.
The large-$k$ behavior of the momentum distribution is given by $n_k \sim \frac{\sigma}{\Gamma_k} \sim k^{-2}$, which is thus determined by the momentum dependence of the polariton linewidth.

The behavior of the generalized Gross Pitaevskii equation in a dynamically unstable regime has been studied in various works \cite{Bobrovska2014, Bobrovska2015,Bobrovska2017, Baboux2018}. While the onset of instabilities can be described in terms of the linearized evolution of Bogoliubov modes, their full development breaks down the Bogoliubov approximation such that their dynamics can only be accessed numerically. 

\subsection{Numerical methods}

We investigate the dynamics of the driven-dissipative exciton-polariton condensate via the numerical integration of the coupled equations (\ref{eq:sGPE}) and (\ref{eq:reservoir}). We fix the system size to $L=1024 a$. For each time step $dt$, we compute the evolution using a standard semi-implicit split-step method for (\ref{eq:sGPE}) and an explicit fourth order Runge-Kutta method for (\ref{eq:reservoir}), then adding to each point of $\psi(x, t)$ the stochastic term $\sqrt{\sigma\left(\frac{dt}{a}\right)}\left(r_1 + ir_2\right)$ using Euler-Maruyama scheme, with $r_{1,2}$ independent random numbers sampled from a normal distribution $\mathcal{N}(0, 1)$.

The runs are performed according to the following procedure.
We set as initial condition the homogeneous mean-field solution $\psi(x, 0) = \sqrt{\rho_0}$. 
We then run the full stochastic evolution, during which we record the variation of the space-averaged density $\bar{\rho}(t) = \frac{1}{L} \int_0^L |\psi(x, t)|^2 dx$, up to a finite time $t_0$ after which the system reaches a steady-state.
The value of $t_0$ depends on the specific parameters of Eqs.~(\ref{eq:sGPE}, \ref{eq:reservoir}) and is thus determined case by case. 
We identify different regimes by characterizing the steady state dynamics of $\psi(x,t)$ and its first-order correlations, where the reference time is set to $t_0$.
We employ the numerical simulations  to explore the different dynamical regimes realized by  varying the value of the blueshift $\mu_{\rm th}$, the reduced pump $p$ and the intensity of stochastic fluctuations $\sigma$.

\section{Results}
\label{sec:results}

\subsection{Kardar-Parisi-Zhang regime}
\label{subsec:kpz_regime}

\begin{figure}[h]
    \includegraphics[width=.99\linewidth]{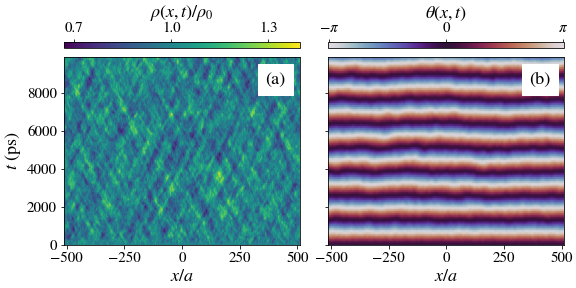}
    \includegraphics[width=.99\linewidth]{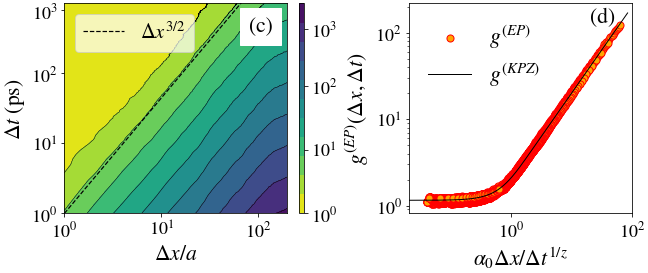}
    \caption{(Color online) (a, b) Space-time maps for the condensate density (in units of $\rho_0$) and for the rescaled phase $\theta(x,t) = \,{\rm Arg}\left[\psi(x,t) \rm{e}^{i\omega_r t}\right]$ in the KPZ regime.
    The values of the parameters are $\mu_{\rm th}=0.1$ \si{meV}, $p=1.15$, $\sigma = 0.1 \sigma_0$. $\hbar \omega_r=0.9\mu_{\rm th}$. 
    Bottom: scaling behavior in the KPZ regime. (c) Rescaled correlator $g^{(EP)}$ (defined in text) with $\beta=1/3$ and $C_0=0.00018$. The KPZ dynamical scaling exponent $z=3/2$ can be read off  from the slope of the iso-correlation lines $\Delta t\sim \Delta x^{z}$. (d) Collapse of the data onto the KPZ universal scaling function $g^{(\rm KPZ)}$. The non-universal constants $C_0$ and $\alpha_0=2.2$ are determined numerically to fit $g^{(\rm KPZ)}$. 
     The parameters are $\mu=0.1$ \si{meV}, $p = 1.15$, $\sigma=0.1 \sigma_0$. The data are obtained by averaging over 1000 independent noise realizations.}
    \label{fig:regime_kpz}
\end{figure}

We start by characterizing the condensate in the parameter region which is dynamically stable according to the Bogoliubov analysis (see  Fig.~\ref{fig:bogo_phasediagram}). 
Typical space-time maps for the condensate density and phase are shown in Fig. \ref{fig:regime_kpz}a,b. In order to highlight at best the features of the dynamics, in the space-time maps the phase is rescaled as $\theta(x, t) = {\rm Arg}\left[ \psi \, e^{i \omega_r t}\right]$, which amounts to shifting the origin of the energies by an arbitrary value $\hbar \omega_r$, suitably chosen to highlight the phase structure.
The images show a relatively smooth density profile, with small modulations due to the noise fluctuations, and a uniform phase front showing no dislocations. 
We then calculate the space-time first-order correlation function $g^{(1)}(\Delta x, \Delta t)$ by averaging over the trajectories obtained for independent noise realizations. The results are shown in Fig. \ref{fig:regime_kpz}c,d. The KPZ universal scaling clearly emerges in the $g^{(1)}$ correlations in the space-time plane when plotting the rescaled correlator $g^{(\rm EP)}(\Delta x,\Delta t) =-2 C_0^{-1} \Delta t^{-2\beta} \log |g^{(1)}(\Delta x,\Delta t)|$, with $\beta=1/3$ the KPZ critical exponent and $C_0$ a normalisation constant. 
Indeed, owing to the expression \eqref{eq:_Ckpz_scalingform},  $g^{(\rm EP)}$ is expected to depend not on space and time independently, but only on the ratio (the scaling variable) $\Delta x/\Delta t^{1/z}$. This is manifest in the density plot as parallel level lines  of slope $\Delta t \sim \Delta x^{3/2}$, and is also confirmed by the data collapse onto a single curve given by the  exact KPZ scaling function $g^{(\rm KPZ)}$ \cite{Prhofer2004}.

\subsection{Soliton-patterned regime}
\label{subsec:solitons} 

We  now analyze  the dynamics of the exciton-polariton condensate in the parameter region which is dynamically unstable according to the linear Bogoliubov analysis (see Fig.~\ref{fig:bogo_phasediagram}), focusing on the low-noise regime $\sigma\le \sigma_0$. We find that the dynamical instability found in the linearized equations of motion within the Bogoliubov approximation favors the emergence of spatial modulations that are then hampered at the nonlinear level, leading to a  patterned regime with a stationary average density lower than in the KPZ regime.

\begin{figure}
    \includegraphics[width=.49\textwidth]{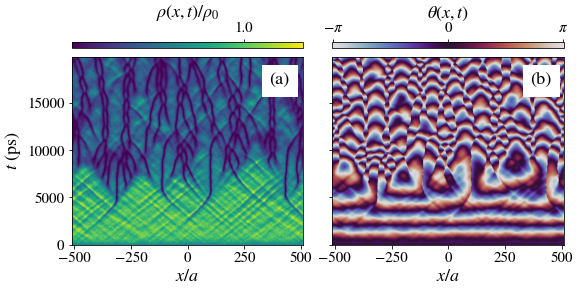}
    \includegraphics[width=.49\textwidth]{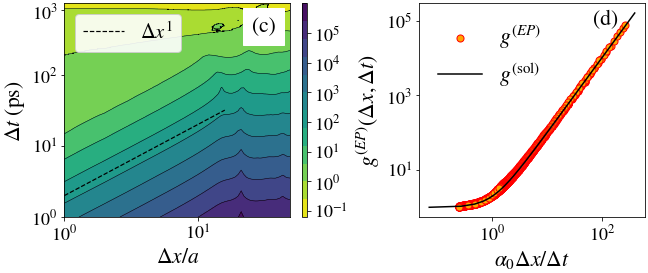}
    \caption{(Color online)
    (a, b)
    Space-time maps for the condensate density (in units of $\rho_0$) and for the rescaled phase $\theta(x,t) = \,{\rm Arg}\left[\psi(x,t) \,e^{i\omega_rt}\right]$ in the soliton-patterned regime.
    The parameters are $\mu_{\rm th}=0.29$ \si{meV}, $p=1.25$, $\sigma = 0.1 \sigma_0$. $\hbar \omega_r=-10\mu_{\rm th}$.
    (c) 
    Rescaled correlator $g^{(EP)}$ (defined in text) with $\beta=1$ and $C_0 = 4.05 \cdot 10^{-5}$. The plot is cut at $\Delta x/a=100$.
    (d)
    collapse of the data, achieved with $\chi=1$ and $z=1$. The function $g^{(\rm sol)}(y)=1 + y^2$, is plotted as a guide, with fitting parameters $C_0$ and $\alpha_0 = 24.15$. 
    The parameters are $\mu=0.6$ \si{meV}, $p = 1.5$, $\sigma=0.1 \sigma_0$.}
    \label{fig:regime_solitons}
\end{figure}

The space-time maps for the density and phase of a typical trajectory are shown in Fig.~\ref{fig:regime_solitons}. At difference with the KPZ regime,  we observe the nucleation and proliferation of long-lived, spatially localized structures characterized by density dips and $\approx \pi$ phase jumps, which dominate the long-time dynamics. We  refer to them as dark solitons, in analogy to the corresponding structures which emerge in the solution of the non-linear Schr\"odinger equation in closed quantum systems. 

\begin{figure}[h]
     \centering
        \includegraphics[width=.47\textwidth]{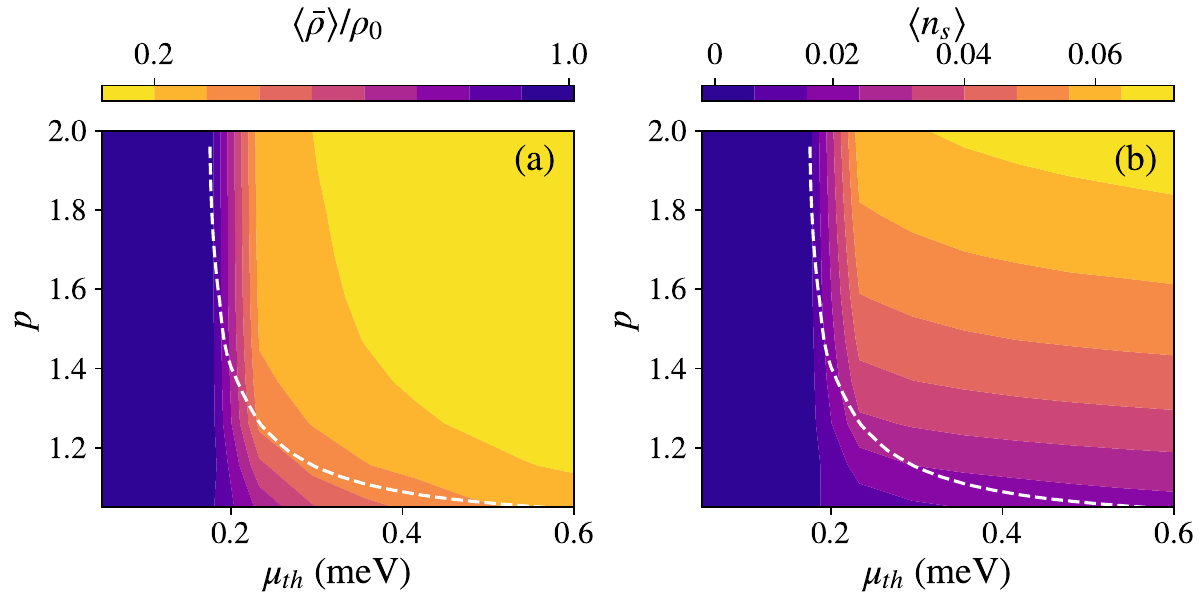}
        \includegraphics[width=.49\textwidth]{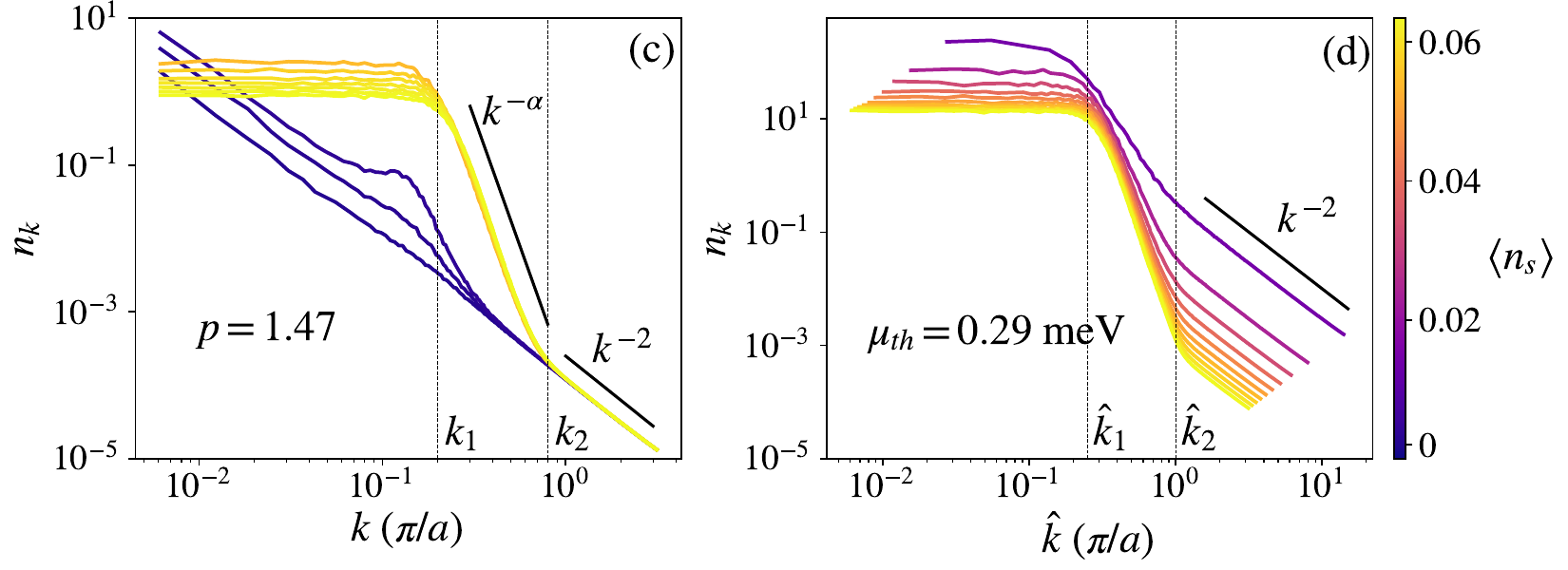}
     \caption{(Color online)
     (a) Average density of the condensate and (b) density of solitons (right panel) in the $(\mu_{\rm th}, p)$ plane (note the inversion of the color scheme).
     The results are obtained by averaging over 100 independent trajectories  starting from a sampling time  $t = 50 000$ $ps$, sufficient to exceed the time needed for the instability to be activated, when present, and for the soliton-patterned regime to fully develop, and then averaged over the subsequent time interval  $\Delta t=5000$ $ps$. The noise is $\sigma = 0.1 \sigma_0$. The dashed line corresponds to the limit of modulational instability determined within the Bogoliubov analysis (see Fig. \ref{fig:bogo_phasediagram}).
     Bottom:
     momentum distribution $n(k)$ (in units of $\rho_0$) as a function of the momentum $k$ (in units of $\rho_0$), for (c) fixed $p$ and (d) fixed $\mu_{\rm th}$. In (d) the momenta are rescaled with the pump as $\hat{k}=k / \sqrt{p-1}$. 
     }
     \label{fig:transition_solitons}
\end{figure}

In order to determine the parameter region where solitons emerge,
we follow the space and time average of the density profile $\langle \bar{\rho} \rangle$ and of the soliton density $\langle n_s \rangle$, shown in Fig. \ref{fig:transition_solitons}. The soliton regime, with $\langle n_s \rangle>0$, is characterized by a strongly reduced density. The information displayed by both indicators is consistent, and globally reflects the picture obtained from the Bogoliubov stability analysis. The Bogoliubov stability diagram thus provides a qualitative determination of the phase boundary of the soliton regime in the $(\mu_{\rm th},p)$ plane. 

Yet, the data from direct numerical simulations exhibits two additional features: i) the soliton-free region is reduced at low pump values compared to the Bogoliubov instability boundary; ii) the density  of dark solitons is strongly dependent on the pump power. This feature was also reported in \cite{Bobrovska2014}. The effect i) is due to a renormalization by the pump power of the relative noise amplitude as $\sigma / |\psi|^2 \propto \sigma/(p-1)$, which becomes large at low pump powers, undermining the validity of the Bogoliubov approximation. This  increase of the effective noise amplitude at small $p$ promotes the random activation of soliton-like defects, which thus leads to a larger density of solitons.
The second feature ii) is connected to the appearance, together with solitons, of  two typical length scales $d_s$ and $L_s$, which can be interpreted as the typical spacing and size of solitons respectively. They are both found to depend on the pump as $d_s, L_s \sim 1 / \sqrt{p-1}$ (see Appendix \ref{appendix:solitons}). They thus affect the total density of solitons which can approximately be estimated as $n_s = 1 / (L_s + d_s) \propto \sqrt{p-1}$. 

The space-time first-order correlation function for the soliton-patterned regime is  shown in Fig.~\ref{fig:regime_solitons}.  
We find that it approximately corresponds to a Gaussian decay in space-time as $g^{(1)}(\Delta x,\Delta t) \sim \mathrm{e}^{-(a\Delta t^2 + b\Delta x^2)}$. The coherence decay hence endows a scaling form,  with the scaling variable  $y=y_0 x/t^{1/z}$ and exponents $\chi=\beta=1$, and thus a dynamical exponent $z=1$,  yielding  a data collapse according to 
\begin{equation}
   -2\mathrm{log}| g^{(1)}(\Delta x,\Delta t)| = C_0^{(sol)} \Delta t^{2} g^{(\rm sol)}\left(\alpha_0^{(sol)} \frac{\Delta x}{\Delta t}\right) 
    \label{eq:scaling_solitons}
\end{equation}
with $g^{(\rm sol)}(y) = 1+y^2$ and $C_0^{(sol)}, \alpha_0^{(sol)}$ normalization constants.
We remark however that the window of spatial and temporal coherence of the condensate is smaller than the one in the KPZ regime (notice the different horizontal scale as compared to Fig. \ref{fig:regime_kpz}). The spatial coherence is limited by the typical spacing between solitons $d_s$.
Let us mention that the scaling $z=1$ was also recently observed in studies of the inviscid, or tensionless, limit of the KPZ equation \cite{Brachet2022,Cuerno2022,Fontaine2023}. The potential connection between the two systems is intriguing and would deserve further theoretical investigation. 

The presence of solitons is also clearly detectable in the momentum distribution $n(k)$, defined by Eq. (\ref{eq:nk}), which is displayed in Fig. \ref{fig:transition_solitons}. One observes that  two scales appear. They  can be identified as $k_1\sim d_s^{-1}$ and $k_2 \sim L_s^{-1}$, the typical spacing and size of solitons.
For $k<k_1$, the momentum distribution is $k$-independent, due to the loss of long-range first-order coherence. 
 In the region $k_1 < k < k_2$, we observe a power law decay $n_k\sim k^{-\alpha}$ with   $\alpha$ a  non-universal exponent in the range $6\lesssim \alpha \lesssim 7$. Similar decays were reported in \cite{Bobrovska2015, He2017}. This behaviour is well marked at high pump power. 
 Finally, 
for $k>k_2$ the momentum distribution  exhibits a power law decay $n_k \sim k^{-2}$. 
This decay is determined by the microscopic details of the model (\ref{eq:sGPE}), and in particular can be related to 
 the quadratic growth of the polaritons linewidth $\gamma(k) = \gamma_0 + \gamma_2 k^2$. 
 It thus has a  very different origin from the one predicted by  \cite{Schmidt2012} in the context of non-thermal Bose gases. 

To further characterize the scaling range where the momentum distribution displays the  $k^{-\alpha}$ decay, we show in Fig. \ref{fig:transition_solitons} the data of $n(k)$ for different $\mu_{\rm th}$ at fixed $p$ (left panel) and for different $p$  at fixed $\mu_{\rm th}$ (right panel). In the latter,  the momentum is rescaled as $k\rightarrow \hat{k} = k / \sqrt{p-1}$, which yields a very good horizontal collapse of the curves at intermediate wavevectors. This shows that the two scales $k_1$ and $k_2$ are independent of $\mu_{\rm th}$, and depend on the pump power with as $\propto 1/\sqrt{p-1}$. The appearance of the power-law scaling $n(k)\sim k^{-\alpha}$ with a large power-law exponent $6\le \alpha \le 7 $ is a clear and robust signature of the soliton regime, which can be used to detect this phase.

\subsection{Disordered vortex regime}
\label{subsec:vortices}

\begin{figure}[h]
    \includegraphics[width=.49\textwidth]{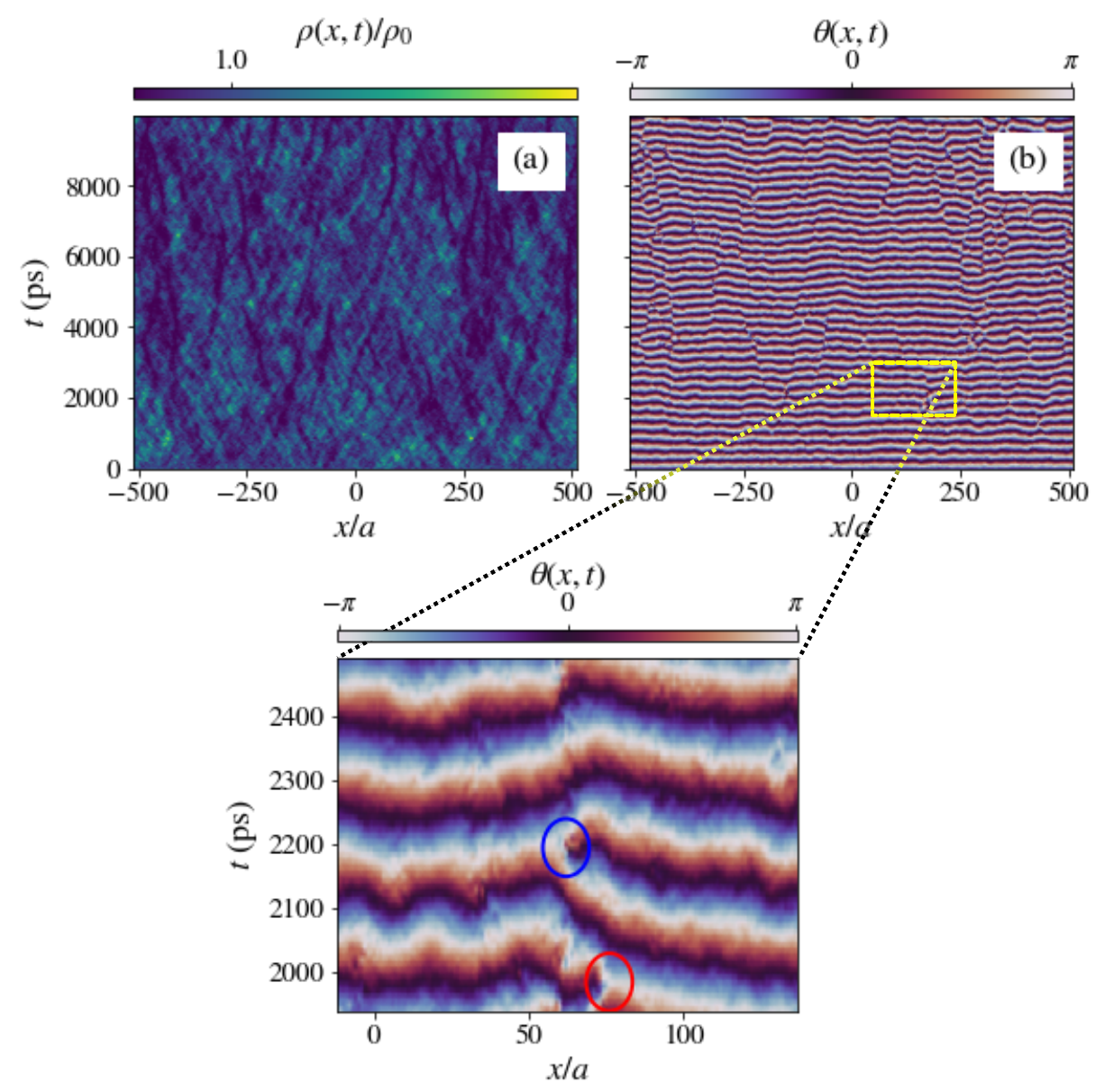}
    \caption{(Color online) (a, b) Space-time maps for the condensate density (in units of $\rho_0$) and for the rescaled phase $\theta(x,t) = \,{\rm Arg}\left[\psi(x,t) \,e^{i\omega_rt}\right]$ in the disordered-vortex regime.
    The parameters are $\mu_{\rm th}=0.1$ \si{meV}, $p=1.15$, $\sigma = 5 \sigma_0$, $\omega_r=0$. 
    In the zoomed window, two spatio-temporal vortices of opposite charge are identified.}
    \label{fig:regime_vortices}
\end{figure}

We now focus on the effect of increasing the noise strength 
starting from the soliton-free region of the phase diagram where the system exhibits KPZ universal features. We show in Fig.~\ref{fig:regime_vortices} a typical space-time map obtained at large noise $\sigma=5\sigma_0$.  In this case the phase map displays a high
density of phase slips, or space-time vortices, appearing as fork-like in the $(x,t)$ plane. 

\begin{figure}
  \includegraphics[width=.99 \linewidth]{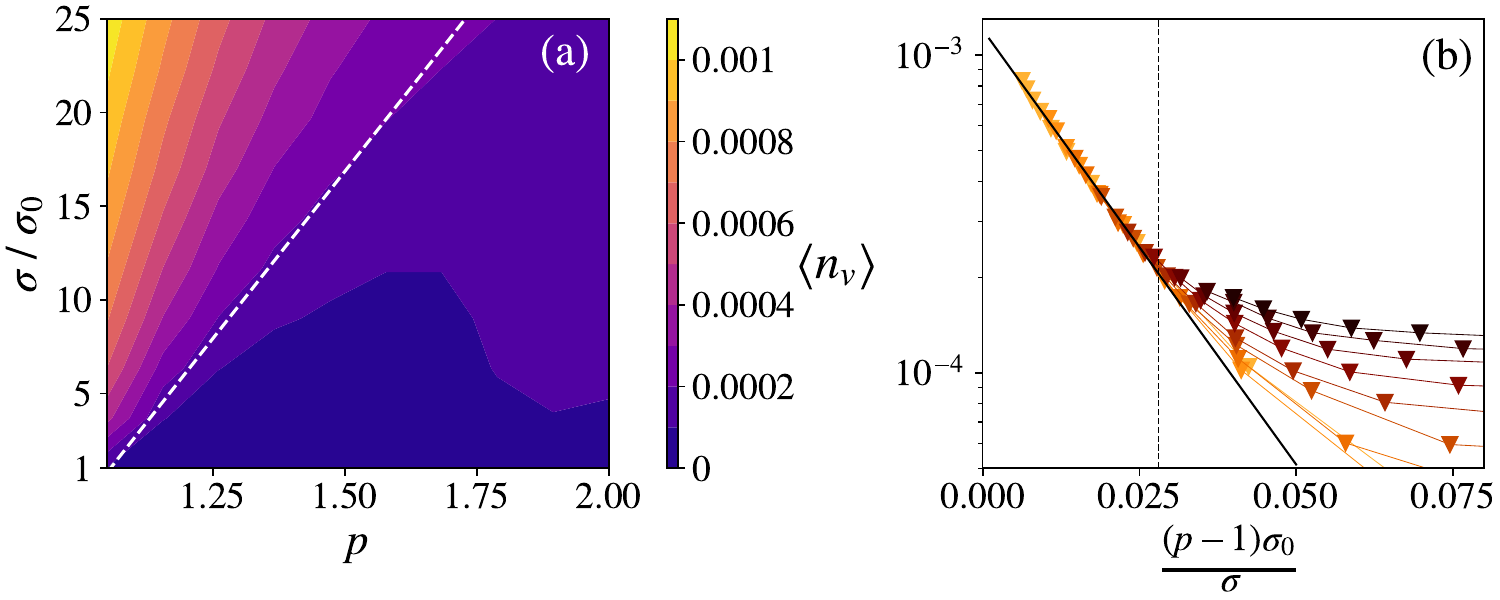}
  \caption{(Color online) (a) Vortex density $\langle n_v\rangle$ in the $(p,\sigma)$ plane. (b) Data for $\langle n_v \rangle$ plotted as a function of $(p-1) \sigma_0/\sigma$, to  evidence its dependence in this combination. The colors represent fixed pump values (increasing from light to dark). The solid line is the exponential fit $\langle n_v\rangle = n_0 \, e^{-A \frac{(p-1) \sigma_0}{\sigma}}$, with $n_0 = 0.012$, $A = 63$.
  A dashed line is plotted in both panels at the limit of validity of the collapse on this exponential behavior, here $(p-1) \sigma_0/\sigma=0.028$.
  }
    \label{fig:transition_vortices}
\end{figure}

In order to characterize the activation of these defects, we investigate the behavior of the defect density $\langle n_v \rangle$ when varying the noise strength and  the pump power. The results are shown in Fig.~\ref{fig:transition_vortices} as a function of the rescaled noise $\sigma / \rho_0 \propto \sigma / (p-1)$. Indeed, this specific combination enters the effective  noise $D$ in the KPZ equation (\ref{eq:kpz_phase}) and hence is the relevant parameter for the stochastic dynamics of the phase.
For high rescaled noise values, the density of vortices displays an exponential behavior $\langle n_v \rangle \propto  \mathrm{exp}\left[-\frac{A  (p-1) \sigma_0}{\sigma}\right]$, where $A$ is a constant which depends on the other parameters of the model, consistently with the results of Ref.~\cite{He2017} where noise-activated space-time vortices were first reported.

\begin{figure}
    \centering\includegraphics[width=.99\linewidth]{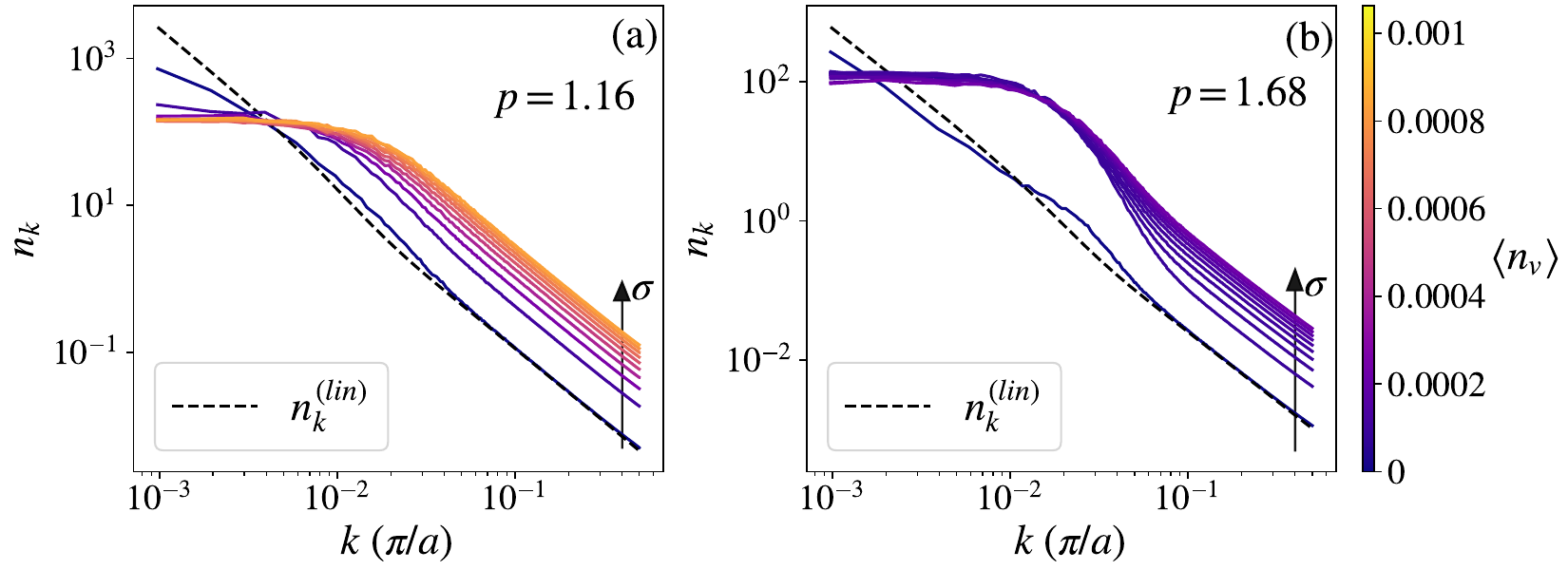}
    \caption{(Color online) Comparison of the behavior of the momentum distributions at increasing noise ($1 \le \sigma \le 25$) for (a) $p=1.16$ and (b) $p=1.68$. The color code indicates the density of vortex-like defects $\langle n_v \rangle$ of Fig. \ref{fig:transition_vortices}.}
    \label{fig:nk_vortices}a
\end{figure}

This feature is progressively lost upon decreasing the noise. Below a certain value, indicated with a dashed line in both left and right panels of Fig.~\ref{fig:transition_vortices},  the noise and pump effects decouple and $\langle n_v \rangle$ departs from the exponential law. We observe that as the pump is further increased, the density of noise-activated defects stabilizes, the density dips lasting longer in time, and  the defects progressively deforming into longer-lived structures.  Their shape is somewhat between vortices and solitons. This feature is also visible in the momentum distribution, as shown in Fig.  ~\ref{fig:nk_vortices}. At low pump, the activation of defects at increased noise translates into  plateaus at low momenta, corresponding to a coherence loss beyond the average vortex separation length. At higher pump, we observe a modulation at intermediate momenta, signature of long-lived defects.

\begin{figure}
    \centering
    \includegraphics[width=.99\linewidth]{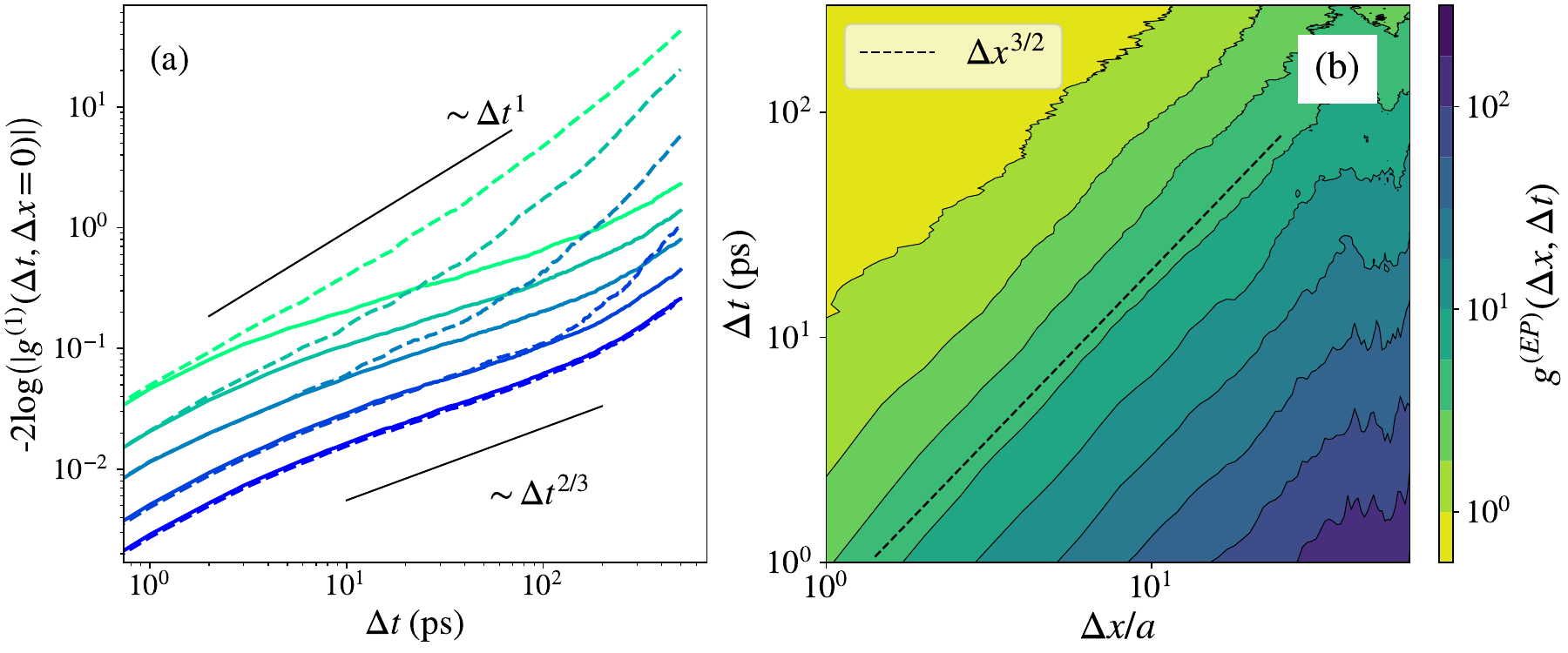}
    \caption{(Color online) (a) Comparison of the temporal correlations of the phase for low pump $p=1.16$ at different noise values (from dark to light blue $\sigma / \sigma_0 = 6, 10, 18, 25, 50$), computed from the coherence $-2\mathrm{log}|g^{(1)}(t, x=0)|$ (solid lines) and directly as $C_{\theta}$ (dashed lines). 
    (b)
    Spatio-temporal coherence decay, compensated by the KPZ scaling as in Fig. \ref{fig:regime_kpz}, for $p=1.15$,  $\sigma = 25 \sigma_0$. The normalisation is $C_0 = 0.045$.}
    \label{fig:transition_vortices_correlations}
\end{figure}

We then determine the first-order correlation function in order to characterize the space-time coherence of the exciton-polariton condensate in the vortex phase. In the presence of topological defects, the validity of the relation (\ref{eq:g1_approx_Ctheta}) connecting the coherence of the condensate $g^{(1)}(\Delta x, \Delta t)$ to the phase correlations $C_{\theta}(\Delta x, \Delta t)$ eventually fails. Indeed, in $g^{(1)}(\Delta x, \Delta t)$ the phase differences only enter in the complex exponential, hence the information about $\approx 2\pi$ phase jumps is lost, whereas in the unwrapped phase correlations, these phase jumps have a strong effect and  dominate the scaling. This feature is visible when the temporal coherence of $g^{(\rm EP)}(\Delta x,\Delta t) =-2  \Delta t^{-2\beta} \log |g^{(1)}(\Delta x,\Delta t)|$ and of the phase-phase correlations are compared for a low value of the pump, as shown in the left panel of Fig.~\ref{fig:transition_vortices_correlations}. The two correlation functions differ for time intervals larger than $t_c$, such that
$C_{\theta}(\Delta t>t_c) > -2\log|g^{(1)}|(\Delta t>t_c)$. Thus at low pump, as a result of this resilience of the coherence to $2\pi$ phase jumps, the KPZ scaling in the coherence persists in the presence of vortices. This is due to the fact that the phase dynamics is piece-wise KPZ, i.e.  the phase fluctuations grow according to the KPZ dynamics within two defects. 

However, what distinguishes the vortex-disordered regime from the KPZ one is that, in the vortex regime, the phase-phase correlations depart from the KPZ universal behavior, and exhibit instead a linear scaling in time, compatible with the effect of random phase jumps, as suggested in Ref.~\cite{He2017}, and similarly to Ref.~\cite{Lauter2017}. When the pump is increased, the reservoir dynamics affects the correlations as is detailed in Sec.\ref{subsec:reservoir_dynamics}.

We note that, due to the phase jumps occurring in both cases, the counting of solitons gives equivalent results as the total vorticity shown in Fig ~\ref{fig:transition_vortices}. Hence, the density of defects alone is not enough to discriminate between the vortex and the soliton phases. The joint analysis of the spatio-temporal correlations is needed, and shows two distinct scaling behaviors in the vortex and soliton phases. 
 
Let us make a final remark concerning the crossover from the KPZ phase to the vortex-disordered  phase. We observed that the density of vortices is a clear marker of the crossover, while the coherence is not very sensitive to it. 
This situation is reminiscent of spatio-temporal intermittency encountered and characterized in the context of the complex Ginzburg-Landau equation \cite{Chate1994, Chat1995, Aranson2002, GarcaMorales2012}.  This regime is usually reported in regions of the parameter-space for which  linear wave solutions are stable.
It is characterized by extended regions of coherent waves which coexist with localized defects, separating them. As  was pointed out in Ref.~\cite{Chat1995}, spatio-temporal intermittency is badly characterized by global observables (such as the average density) or locally probed averages (such as space-time correlations via $g^{(1)}$). One rather needs to follow the full evolution of an observable in order to identify the actual dynamical regime. It is indeed the unwinding of the phase along the time axis that allows one to detect the crossover in the phase-phase correlations $C_\theta$.

\subsection{Reservoir-textured regime}
\label{subsec:reservoir_dynamics}

\begin{figure}
    \centering\includegraphics[width=.99 \linewidth]{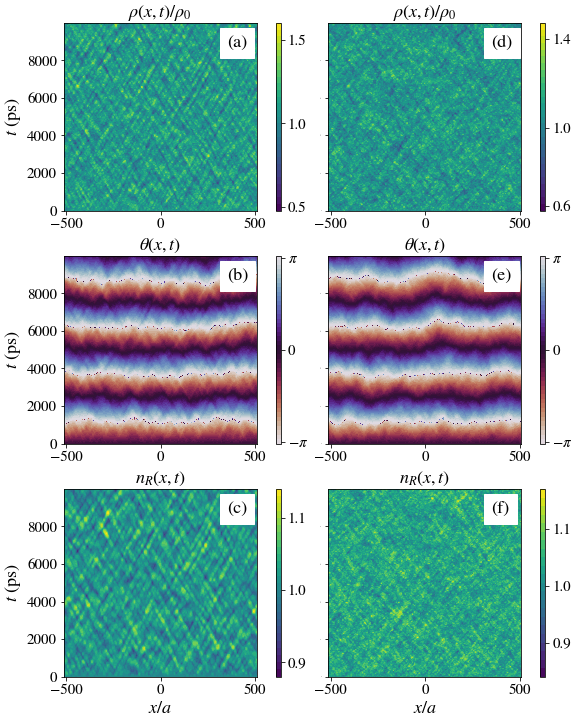}
    \caption{(Color online) (a,b,c) Simulated evolution of the condensate density $\rho$ (units of $\rho_0$), phase $\theta = {\rm Arg} \left[\psi \, e^{i \omega_r t}\right]$ and reservoir density $n_R$ (in units of $n_{R0} = \gamma_0/R$), compared with (d,e,f) a simulation with the same parameters but implementing the adiabatic approximation for the reservoir. The parameters are $p=1.68$, $\sigma=\sigma_0$, $\mu_{\rm th}=0.1$ \si{meV}, $\hbar \omega_r = 0.1 \mu_{\rm th}$.}
    \label{fig:adiab-nonadiab-chart_comparison}
\end{figure}

We finally investigate the mechanism responsible for the departure from the KPZ regime at high pump, focusing on the case $\sigma= \sigma_0$, where topological defects are rare. In this case, the fading of the KPZ scaling is not associated with the appearance of defects, but can be attributed to the fact that the coupling between the reservoir and the condensate dynamics cannot be neglected, and the adiabatic approximation does not hold any more. To show this, one can compare the results obtained by running the full model \eqref{eq:sGPE} and \eqref{eq:reservoir}, and by running its adiabatic approximation, where the reservoir dynamics is replaced by its stationary solution ($\partial_t n_R=0$ in \eqref{eq:reservoir}). This comparison is shown in Fig. \ref{fig:adiab-nonadiab-chart_comparison}, where
typical space-time maps of the condensate phase and density, and of the reservoir density are displayed in both cases. In the full model, the density fluctuations are larger in the non-adiabatic case as compared to the adiabatic case, forming structures that we call textures. In Fig.\ref{fig:adiab-nonadiab-chart_comparison}.(c) We observe that this texture contrast is increased at higher pump power. This comparison clearly indicates that this texturing results from the coupled condensate and reservoir dynamics.

\begin{figure}
    \centering
    \includegraphics[width=.6 \linewidth]{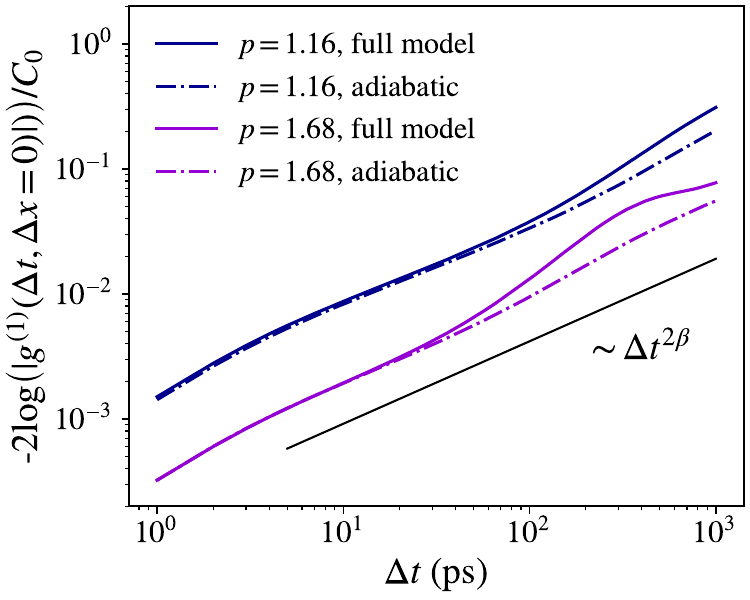}
    \caption{(Color online) Temporal coherence decay for  two values of the pump $p=1.16$ and $p=1.68$ at $\sigma=1.0$, $\mu_{\rm th}=0.1$ \si{meV}, computed from the general case without adiabatic approximation (solid line), {\it i.e} the coupled Eqs. \ref{eq:sGPE}, \ref{eq:reservoir}, and from its adiabatic approximation (dashdot line). The black line serves as a guide to show the KPZ scaling $\beta = 1/3$.}
    \label{fig:adiab-nonadiab-g1_comparison}
\end{figure}

In Fig.\ref{fig:adiab-nonadiab-g1_comparison}, we compare $g^{(1)}(\Delta t,\Delta x=0)$, as calculated from Eqs.(\ref{eq:sGPE}),(\ref{eq:reservoir}) without any approximation, to the adiabatic one for a low and a high pump value. While they are very similar at small pump, they differ at large pump, where a structure appears at large $\Delta t$ in the non-adiabatic case. This structure is absent in the adiabatic approximation where a KPZ scaling persists in the coherence even at large pump.
  
\begin{figure}
    \centering
    \includegraphics[width=.99\linewidth]{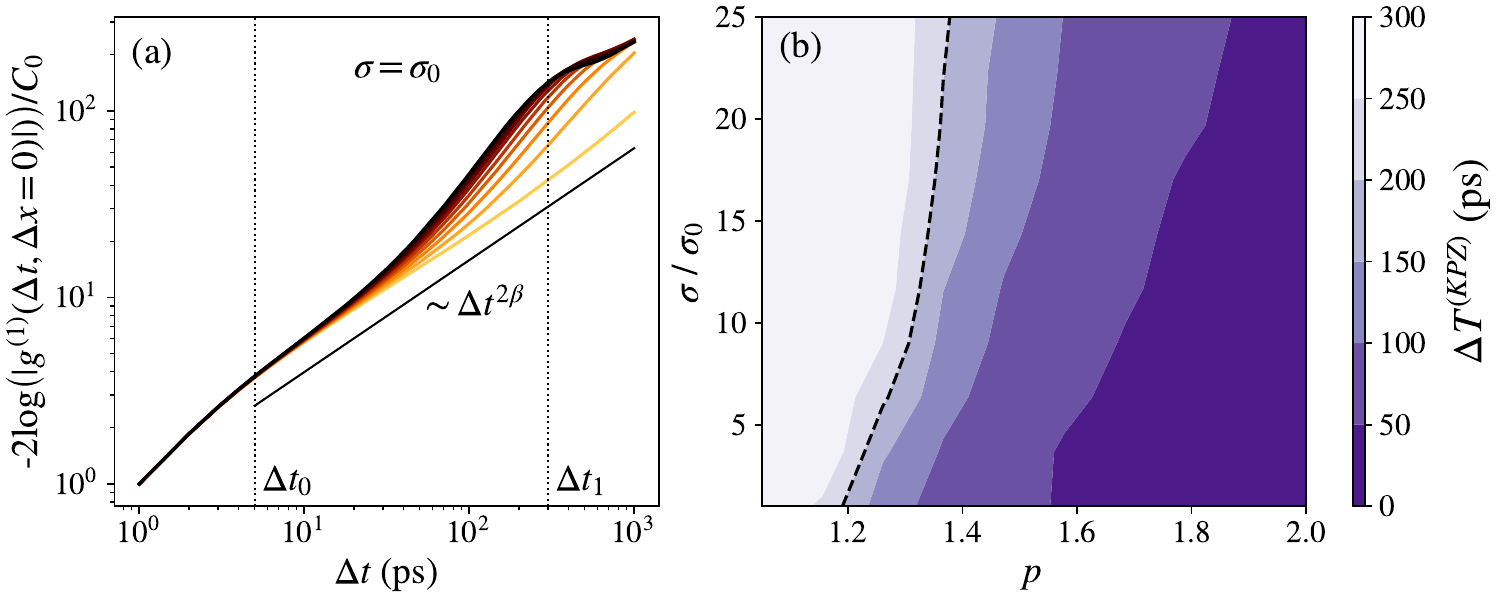}
    \caption{(Color online) (a) Behavior of the temporal coherence decay at different values of the pump (from $p=1.05$, yellow, to $p=2.0$, dark brown ) with $\sigma = 1$. (b) Temporal window of KPZ scaling $\Delta T ^{(\rm KPZ)}$ (see text). The black dotted line corresponds to $\Delta T ^{(\rm KPZ)} = 200$ ps.}
    \label{fig:reservoir_kpz_window}
\end{figure}

Although the phase remains defect-free in this regime (see Fig. \ref{fig:adiab-nonadiab-chart_comparison}), the coherence of the condensate, measured by $-2\log|g^{(1)}|$, progressively looses the KPZ scaling upon increasing pump power, as shown in the left panel of Fig. \ref{fig:reservoir_kpz_window}. 
This comes with a modulation of the phase correlations, resulting in a faster decay of the coherence which deviates from the KPZ decay. This can be quantified by determining the delay window $\Delta T^{(\rm KPZ)}$ in which the KPZ scaling persists. For this, we perform a fit of the power law $\Delta t^{2\beta}$ over increasing time windows $\left[ \Delta t_0 =10, \Delta t_1 \right]$, computing the exponent $\beta$ for each $\Delta t_1$. We stop the iteration and define the KPZ delay window $\Delta T ^{(\rm KPZ)} = \Delta t_{stop} - \Delta t_0$ when $2 \beta > 0.7$. 
The outcome of this procedure in the whole $(p,\sigma)$ plane is displayed in the right panel of Fig. \ref{fig:reservoir_kpz_window}. We observe that the departure from the KPZ phase when increasing the pump intensity is only weakly dependent on the noise strength. We emphasize that a similar observation of the shrinking of the KPZ window as the pump power is increased was reported in Ref.~\cite{Fontaine2022},  both from experimental measurements and numerical simulations.  

In light of this analysis, let us come back to the behavior of the vortex density and momentum distribution in the $(p,\sigma)$ plane discussed in Sec. \ref{subsec:vortices}. In fact, we find that at high pump power, as the reservoir dynamics becomes important, the noise fluctuations can activate sparse long-lived solitons. These structures are responsible for the convergence of the computed vortex density to a small but non-zero value visible in Fig.\ref{fig:transition_vortices} on the right side of the dashed lines, and also to the corresponding modulation in the momentum distribution of Fig.~\ref{fig:nk_vortices} at high pump. We hence attribute these two effects to the role of the reservoir. Finally, let us mention that the space-time maps in this regime are similar to those reported in Ref.~\cite{Bobrovska2014} under the label of non-adiabatic instability, although the model and parameters studied in this reference are different from ours and cannot be directly compared. 

\subsection{Phase Diagram}
\label{subsec:phase_diagram}

\begin{figure}
    \centering
    \includegraphics[width=.65\linewidth]{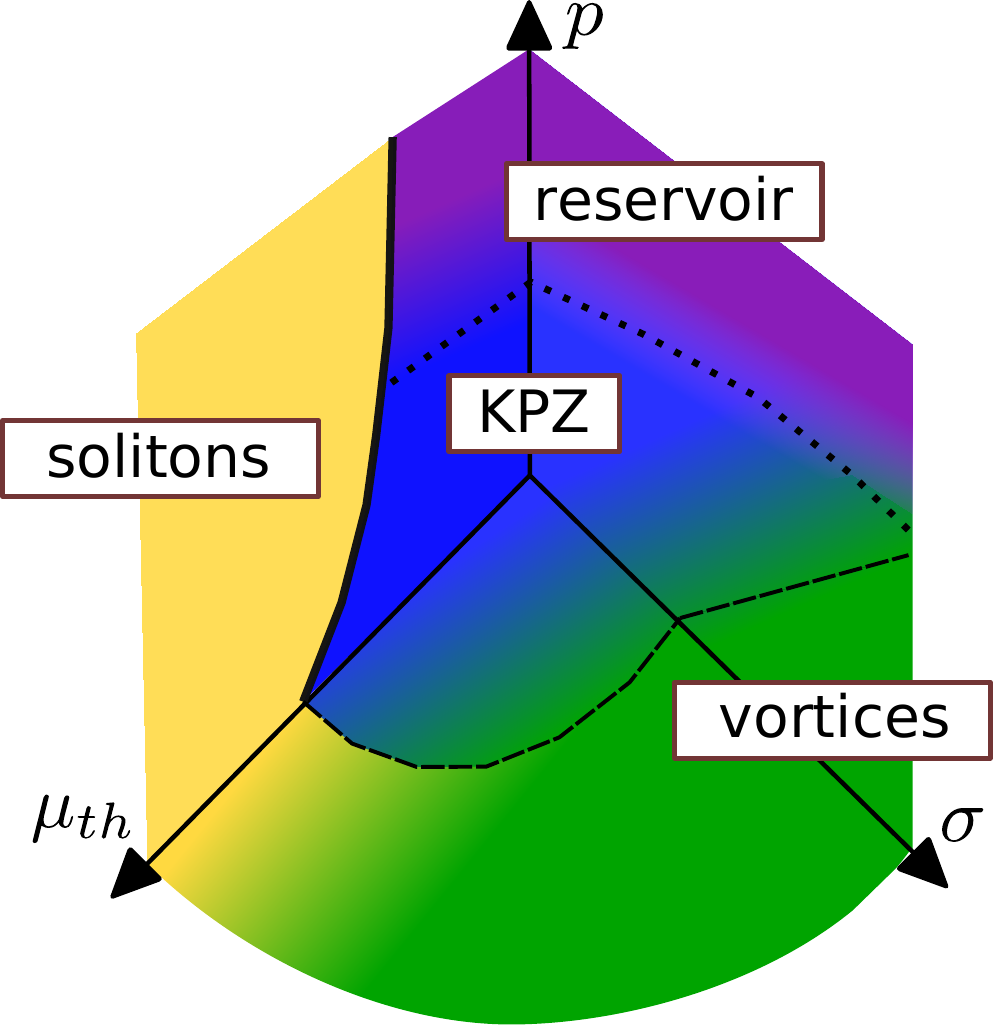}
    \caption{(Color online) Schematic phase diagram of the different regimes, merging the results of Figs.\ref{fig:transition_solitons}.b,
    \ref{fig:transition_vortices}.a and \ref{fig:nv_mu_sigma}. The representation of the four regimes and their transitions is depicted by using four colors and different shadings.
    The $(\mu_{\rm th}, p)$ plane is based on the characterization of Sec. \ref{subsec:solitons} at $\sigma = 0.1 \sigma_0$. The thick solid line represents the limit of modulational instability.
    The $(p, \sigma)$ plane is based on the characterization of Sec. \ref{subsec:vortices} and \ref{subsec:reservoir_dynamics} at $\mu_{\rm th}=0.1$ \si{meV}. 
    The $(\mu_{\rm th}, \sigma)$ plane is based on the characterization at $p=1.16$ reported in appendix \ref{appendix:mu_sigma}.
    The dashed lines represent the onset of a finite defect density, while the dotted lines represent the ending of the KPZ time window in the first-order correlation function.  
    }
    \label{fig:3d_phase_diagram}
\end{figure}

\begin{figure}
    \centering
    \includegraphics[width=.4\textwidth]{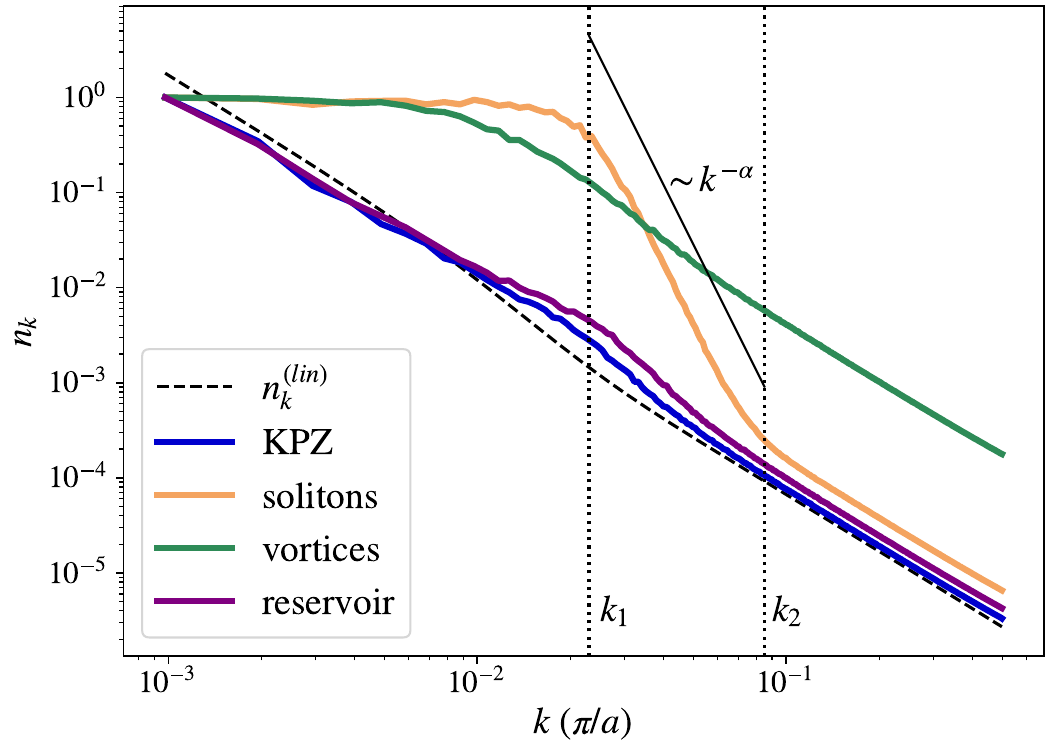}
    \caption{(Color online) Comparison of the momentum distribution $n_k$ in the four regimes. The parameters are: $p=1.26$, $\mu_{\rm th}=0.10$ \si{meV}, $\sigma =0.1$ (KPZ regime); $p=1.26$, $\mu_{\rm th}=0.30$ \si{meV}, $\sigma =0.1$ (soliton regime); $p=1.26$, $\mu_{\rm th}=0.10$ \si{meV}, $\sigma = 9.0$ (vortex regime); $p=1.68$, $\mu_{\rm th}=0.10$ \si{meV}, $\sigma =1.0$ (reservoir regime). The vertical axes are rescaled to facilitate the comparison in the low momentum sector. The prediction of the linear theory is denoted $n_k^{(lin)}$ for the parameters of the KPZ regime.}
   \label{fig:nk_comparison}
\end{figure}

Our analysis is summarized in  Fig.~\ref{fig:3d_phase_diagram},  displaying  a three-dimensional schematic phase diagram at varying pump, noise and interaction strengths, which collects at a glance our results of Figs.~\ref{fig:transition_solitons}.b, \ref{fig:transition_vortices}.a and \ref{fig:nv_mu_sigma}. 
The diagram shows the transition from the KPZ phase to the soliton phase at increasing  $\mu_{\rm th}$, and the crossovers to the vortex-disordered phase at increasing  $\sigma$, and to the reservoir-textured phase at increasing $p$, monitored  by following the soliton and vortex densities, and the size of the KPZ window in time. 

For the experimentally realistic model given by the coupled Eqs.(\ref{eq:sGPE}) and (\ref{eq:reservoir}), we highlight the existence of two new regimes in addition to the vortex one pointed out in \cite{He2017}, respectively characterized by the presence of solitons or  reservoir-induced textures.

We also compare in Fig.~\ref{fig:nk_comparison} the momentum distribution in the four regimes. In the KPZ and reservoir-textured regimes, the momentum distribution scales as $k^{-2}$ at intermediate wavevectors, like in the linear theory obtained from Bogoliubov approximation. Indeed, in one dimension, the exponential decay of the equal-time spatial correlations for KPZ and in the equilibrium (linear) theory, have the same $\chi=1/2$ exponent, which is equivalent to a $k^{-2}$ decay in Fourier space.
The soliton regime is signaled by the clear appearance of the $k^{-\alpha}$ scaling, whereas in the vortex regime, the momentum distribution exhibits a plateau at small  $k<k_1$.

\section{Conclusion}
In this work, we have obtained the phase diagram of a one-dimensional driven-dissipative exciton polariton condensate at varying interaction, noise and pump magnitudes. We have identified three regimes competing with the KPZ one: at strong interactions, a soliton-patterned regime arises from the dynamical instability of the Bogoliubov excitations at linear level; at higher noise, a vortex-disordered regime sets in due to the proliferation of phase slips, which are space-time point-like defects; while at high pump, a defect-free reservoir-textured regime emerges due to the coupling between the reservoir and the condensate dynamics leading to a non-adiabatic behaviour. We have characterized the space-time coherence in all the regimes and identified the momentum distribution as a key observable to detect the soliton regime. The vortex regime is characterized by a linear increase of the phase-phase time correlations, while the first-order correlation function is weakly affected by vortices and still displays the same critical exponents as in the KPZ region. Based on current experimental possibilities, we plan to extend this analysis to two spatial dimensions, where additional regimes have been proposed to compete with the KPZ one \cite{Diessel2022, Dagvadorj2021, Gladilin2017, Grinstein1996, Deligiannis2022, Comaron2021, Ferrier2022, Dagvadorj2015}. Finally, we emphasize that the versatility of the exciton-polariton platform allows tuning a great variety of parameters, such as the drive amplitude, the sample temperature, the pump noise, the material composition, etc. Therefore, navigating through various phases and exploring the rich phase diagram evidenced in this work is experimentally within reach.

\begin{acknowledgments}
FV acknowledges the support from France 2030 ANR QuantForm-UGA. AM acknowledges funding from the Quantum-SOPHA ANR Project ANR-21-CE47-0009. LC acknowledges support from Institut Universitaire de France (IUF). MR acknowledges support by the Centre of Quantum Technologies Exploratory Initiative programme. QF, SR, and JB acknowledge support from the Paris Ile-de-France Region in the framework of DIM QuanTIP, the French RENATECH network, the H2020-FETFLAG project PhoQus (820392), the European Research Council via projects StG ARQADIA (949730) and AdG ANAPOLIS (101054448).
\end{acknowledgments}

\appendix

\section{Details on the soliton-patterned regime}
\label{appendix:solitons}

\subsection{Relevant scales}
\begin{figure}
    \includegraphics[width=.6\linewidth]{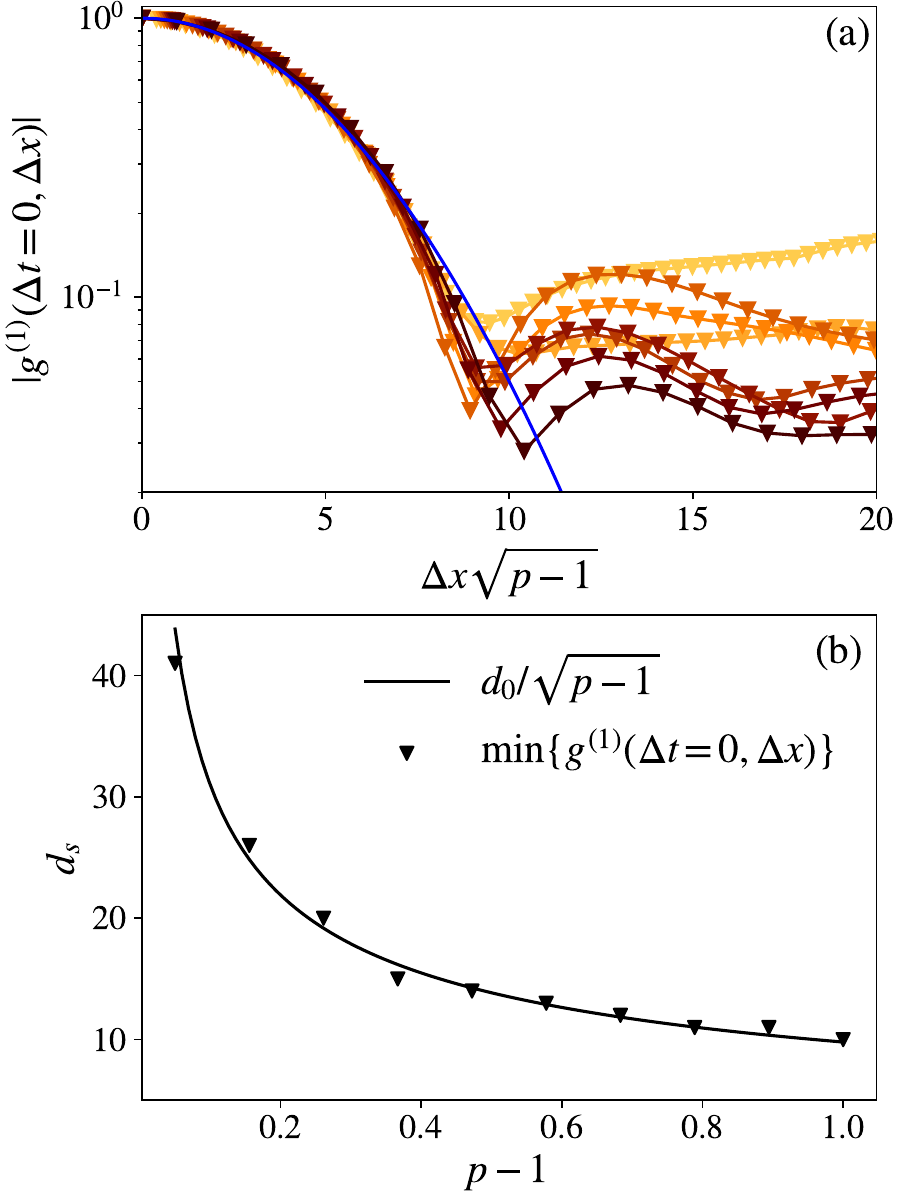}
    \caption{(Color online) (a) Zero-time spatial correlations $g^{(1)}\Delta x, \Delta t=0)$. The Gaussian $e^{-a\Delta x^2}$ with  fitted value $a=0.03$ is plotted in solid black line for comparison. (b)  Typical soliton spacing $d_s$ (in lattice sites), estimated as the minimum of $g^{(1)}(\Delta x, \Delta t=0)$.}
    \label{fig:ds_from_g1}
\end{figure}

In order to estimate the typical soliton spacing $d_s$, we consider the equal-time first-order correlation function $g^{(1)}(\Delta x, \Delta t=0)$. 
Its behavior for different pump powers is shown in Figure \ref{fig:ds_from_g1}, where the space axis has been rescaled as $\Delta x\rightarrow \Delta x\sqrt{p-1}$. 
The Gaussian decay is followed by oscillations which are due to the presence of long-lived soliton patterns. 
We estimate the typical soliton spacing $d_s$ by taking the first minimum of $|g^{(1)}(x, t=0)|$. We find that it depends on the pump as $d_s \sim (p-1)^{-1/2}$, as anticipated in the main text. Moreover, we checked the dependence of $d_s$ on the diffusion constant $\gamma_2$, finding $d_s \sim \sqrt{\gamma_2}$.

\subsection{First-order transition}
\label{appendix:soliton_firstorder_transition}
\begin{figure}
    \includegraphics[width=.23\textwidth]{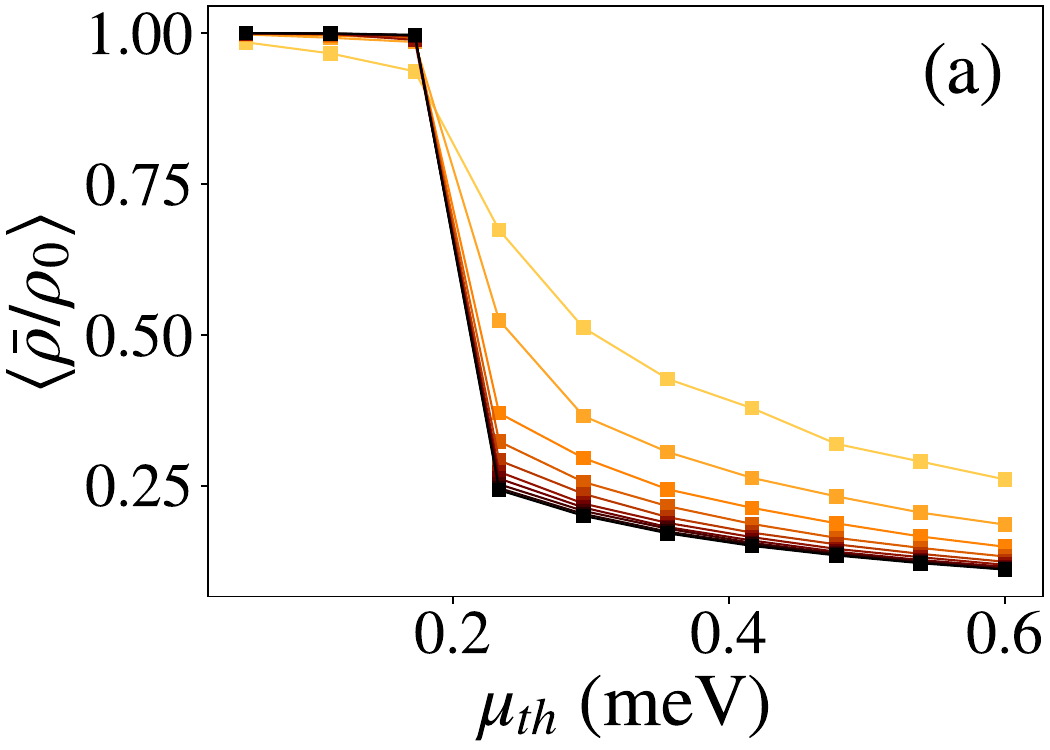}
    \includegraphics[width=.23\textwidth]{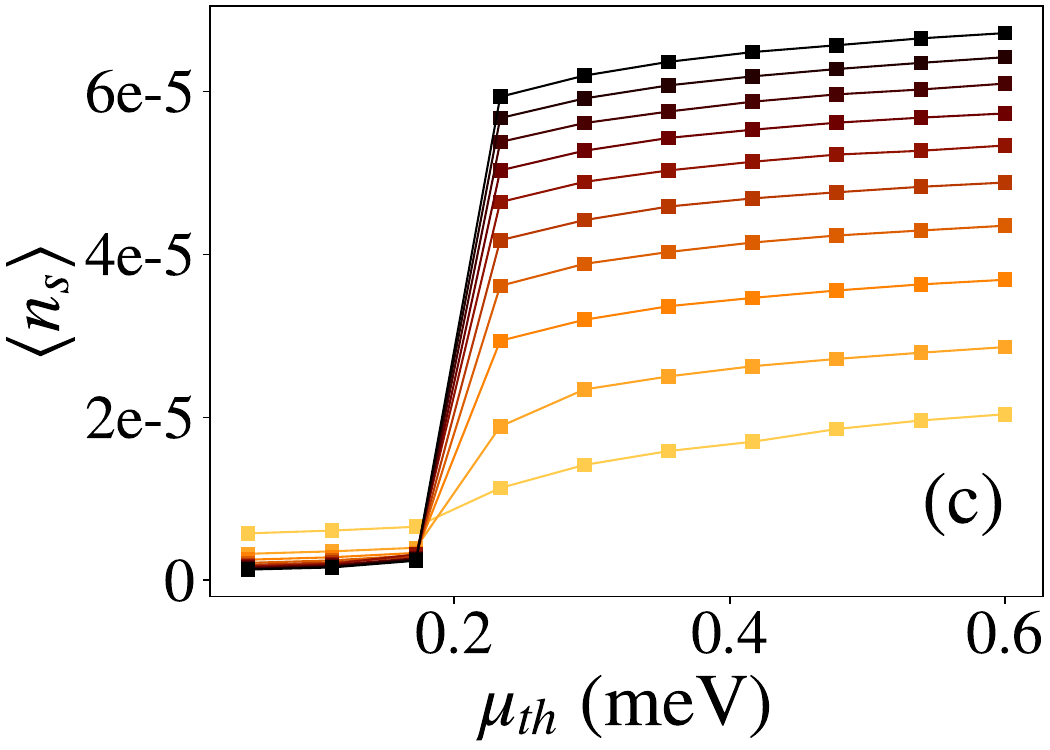}
    \includegraphics[width=.23\textwidth]{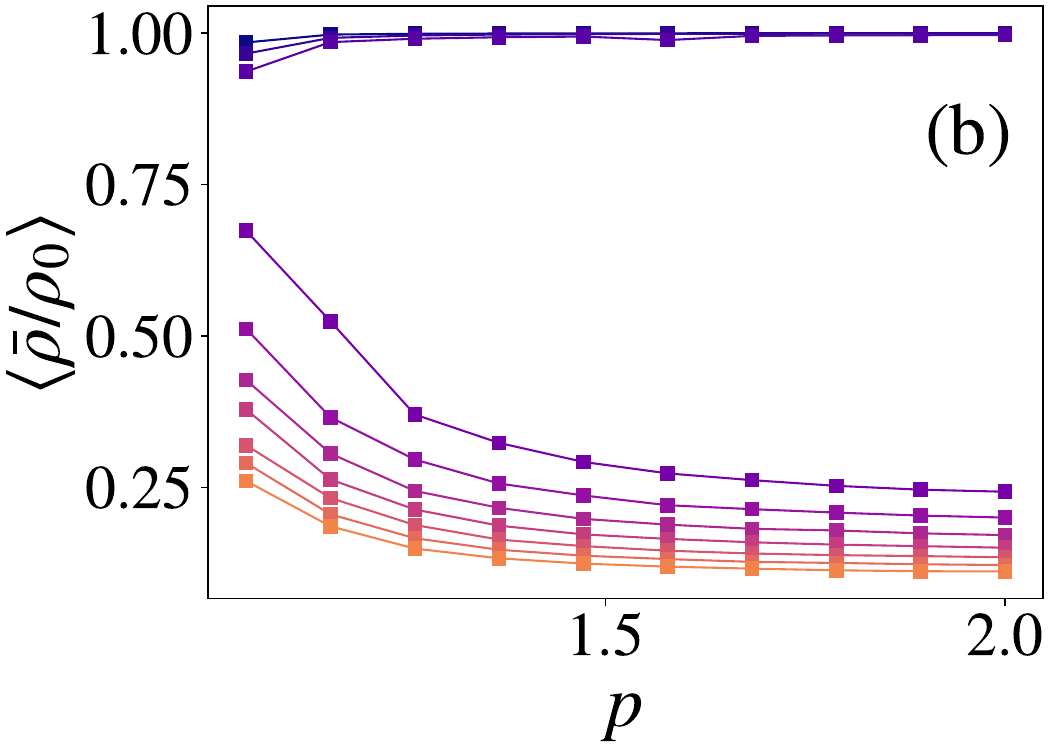}
    \includegraphics[width=.23\textwidth]{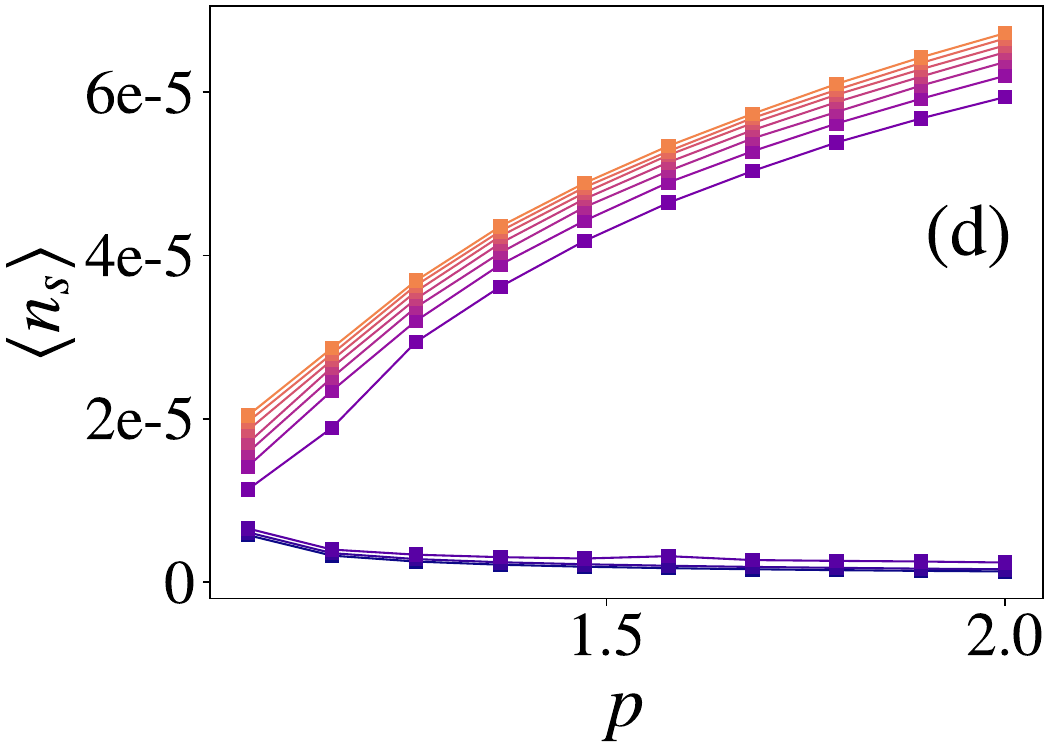}
     \caption{(Color online) Average condensate density (a, b) and average density of solitons (c, d). (a, c) Fixed pump powers, equally spaced from $p=1.05$ (dark) to $p=2.0$ (light).
     (b, d) Fixed interaction strengths, equally spaced from blue $\mu_{\rm th}=0.05$ \si{meV} to rose $\mu_{\rm th} = 0.6$ \si{meV}.}
     \label{fig:app:solitons_mu_detail}
\end{figure}

We present in Fig.~\ref{fig:app:solitons_mu_detail} two cuts of Fig.~\ref{fig:transition_solitons} which shows the average density of the condensate and the average number of solitons: one as a function of interaction strength for different pump powers, and the other as a function of the pump power for different interaction strengths. At increasing interaction strength,  we observe  for all pump powers an abrupt   jump of both $\langle \rho\rangle$ and $\langle n_s\rangle$ at a transition value $\mu_{\rm tr}\sim 0.2$. Similarly, the curves as a function of $p$ split in two clearly distinct sets for $\mu_{\rm th}>\mu_{\rm tr}$ and  $\mu_{\rm th}<\mu_{\rm tr}$. Both observations indicate  a first-order phase transition, as was reported in Ref.~\cite{He2017}. Note that in Ref.~\cite{Dagvadorj2021}, an analogue transition was observed for a two-dimensional exciton-polariton condensate, as a consequence of the  Benjamin-Ferr instability by tuning the pump power below the critical value for the instability.

\subsection{The ($\mu_{\rm th}$, $\sigma$) diagram}
\label{appendix:mu_sigma}
\begin{figure}
    \centering
    \includegraphics[width=.8\linewidth]{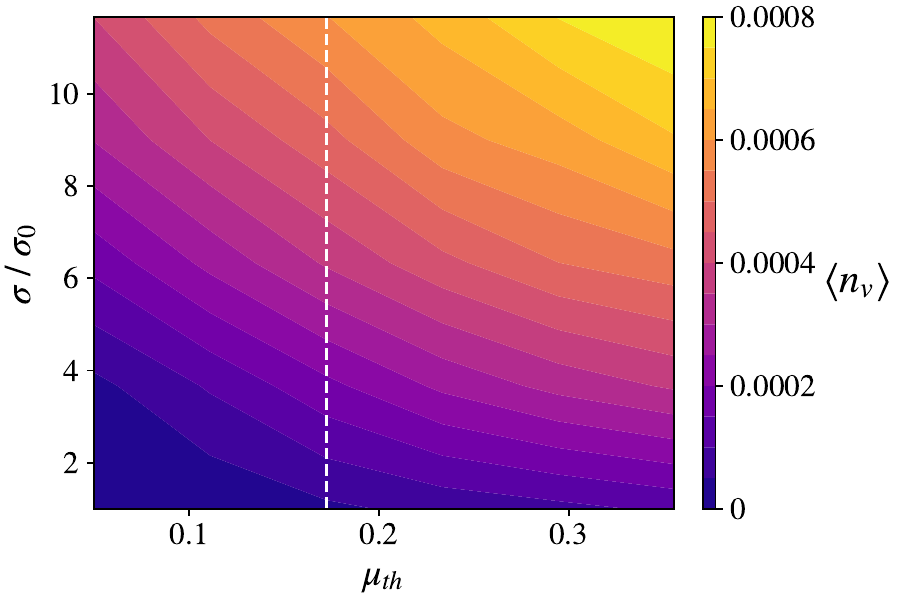}
    \caption{(Color online) Average vortex density in the $(\mu_{\rm th}, \sigma)$ plane, for $p=1.16$. The dashed line indicates the limit of Bogoliubov instability. }
    \label{fig:nv_mu_sigma}
\end{figure}

The density of vortices $\langle n_v\rangle$ computed in the $(\mu_{\rm th}, \sigma)$  for a low pump value $p=1.16$ is reported in Fig.~\ref{fig:nv_mu_sigma}.  The first order transition, shown at low noise in Sec. \ref{subsec:solitons} and Appendix \ref{appendix:soliton_firstorder_transition} is not visible here since 
 the number of noise-activated defects is already non-negligible for the lowest value $\sigma= \sigma_0$ considered for this plot. 
A joint analysis of the momentum distributions  (not shown here) performed at $\sigma= \sigma_0$ has confirmed that  the KPZ to soliton transition is still marked by a change of  behaviour at finite momenta, and in particular 
the appearance of the high-power law decay $n(k)\sim k^{-\alpha}$ upon entering the soliton regime.

\subsection{Nature of the solitons}

\begin{figure}
    \includegraphics[width=.45\textwidth]{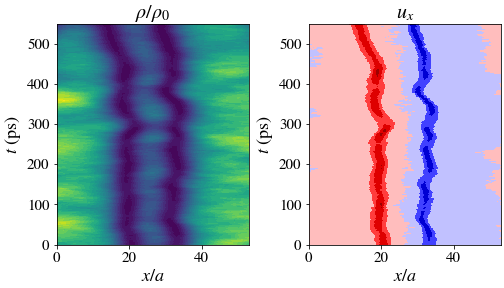}
    \includegraphics[width=.45\textwidth]{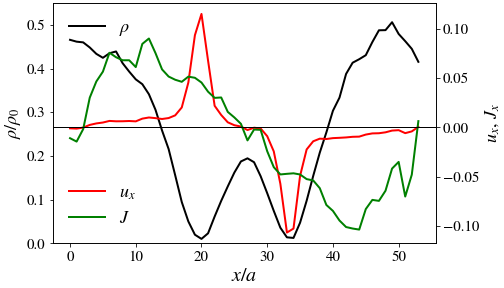}
    \caption{(Color online) Top panels: typical space-time map of a soliton pair and of the associated fluid velocity. Bottom panel: space cut of the density, the velocity, and the current  profiles across a soliton pair, which elucidate its structure.}
    \label{fig:soliton_pair}
\end{figure}

In the one dimensional system studied in Ref.~\cite{Bobrovska2019},  the authors report a chaotic dynamics characterized by nucleation and tree-like branching of defects in space-time. In our case, we do not observe the same branching mechanism. 
The model we consider in this work differs from the one studied in Ref.~\cite{Bobrovska2019} by the sign of the mass (negative here), the interaction energy (reservoir term $2g_R n_R$ only) and the presence of stochastic fluctuations during the whole evolution. It follows that the dynamics is qualitatively different. 
In our case, we rather observe the emergence of metastable soliton pairs, as we show in Fig. \ref{fig:soliton_pair}. 
To explain this feature, we calculate the fluid velocity and current, given respectively by 
\begin{align}
    &u_x = \frac{\hbar}{m} \partial_x \theta \\
    &J = -i \frac{\hbar}{m}\left( \psi^* \partial_x \psi - \partial_x \psi^* \psi \right).
\end{align}
A typical configuration is shown in Fig. \ref{fig:soliton_pair}. 
Both the velocity and the current give a zero total contribution across the  soliton pair, 
 which hence does not split apart. The soliton pair is stable in our case, 
contrarily to  Ref.~\cite{Bobrovska2019}.

\begin{thebibliography}{43}%
\makeatletter
\providecommand \@ifxundefined [1]{%
 \@ifx{#1\undefined}
}%
\providecommand \@ifnum [1]{%
 \ifnum #1\expandafter \@firstoftwo
 \else \expandafter \@secondoftwo
 \fi
}%
\providecommand \@ifx [1]{%
 \ifx #1\expandafter \@firstoftwo
 \else \expandafter \@secondoftwo
 \fi
}%
\providecommand \natexlab [1]{#1}%
\providecommand \enquote  [1]{``#1''}%
\providecommand \bibnamefont  [1]{#1}%
\providecommand \bibfnamefont [1]{#1}%
\providecommand \citenamefont [1]{#1}%
\providecommand \href@noop [0]{\@secondoftwo}%
\providecommand \href [0]{\begingroup \@sanitize@url \@href}%
\providecommand \@href[1]{\@@startlink{#1}\@@href}%
\providecommand \@@href[1]{\endgroup#1\@@endlink}%
\providecommand \@sanitize@url [0]{\catcode `\\12\catcode `\$12\catcode
  `\&12\catcode `\#12\catcode `\^12\catcode `\_12\catcode `\%12\relax}%
\providecommand \@@startlink[1]{}%
\providecommand \@@endlink[0]{}%
\providecommand \url  [0]{\begingroup\@sanitize@url \@url }%
\providecommand \@url [1]{\endgroup\@href {#1}{\urlprefix }}%
\providecommand \urlprefix  [0]{URL }%
\providecommand \Eprint [0]{\href }%
\providecommand \doibase [0]{https://doi.org/}%
\providecommand \selectlanguage [0]{\@gobble}%
\providecommand \bibinfo  [0]{\@secondoftwo}%
\providecommand \bibfield  [0]{\@secondoftwo}%
\providecommand \translation [1]{[#1]}%
\providecommand \BibitemOpen [0]{}%
\providecommand \bibitemStop [0]{}%
\providecommand \bibitemNoStop [0]{.\EOS\space}%
\providecommand \EOS [0]{\spacefactor3000\relax}%
\providecommand \BibitemShut  [1]{\csname bibitem#1\endcsname}%
\let\auto@bib@innerbib\@empty
\bibitem [{\citenamefont {Carusotto}\ and\ \citenamefont
  {Ciuti}(2013)}]{Carusotto2013}%
  \BibitemOpen
  \bibfield  {author} {\bibinfo {author} {\bibfnamefont {I.}~\bibnamefont
  {Carusotto}}\ and\ \bibinfo {author} {\bibfnamefont {C.}~\bibnamefont
  {Ciuti}},\ }\bibfield  {title} {\bibinfo {title} {Quantum fluids of light},\
  }\href {https://doi.org/10.1103/revmodphys.85.299} {\bibfield  {journal}
  {\bibinfo  {journal} {Reviews of Modern Physics}\ }\textbf {\bibinfo {volume}
  {85}},\ \bibinfo {pages} {299} (\bibinfo {year} {2013})}\BibitemShut
  {NoStop}%
\bibitem [{\citenamefont {Weisbuch}\ \emph {et~al.}(1992)\citenamefont
  {Weisbuch}, \citenamefont {Nishioka}, \citenamefont {Ishikawa},\ and\
  \citenamefont {Arakawa}}]{Weisbuch1992}%
  \BibitemOpen
  \bibfield  {author} {\bibinfo {author} {\bibfnamefont {C.}~\bibnamefont
  {Weisbuch}}, \bibinfo {author} {\bibfnamefont {M.}~\bibnamefont {Nishioka}},
  \bibinfo {author} {\bibfnamefont {A.}~\bibnamefont {Ishikawa}},\ and\
  \bibinfo {author} {\bibfnamefont {Y.}~\bibnamefont {Arakawa}},\ }\bibfield
  {title} {\bibinfo {title} {Observation of the coupled exciton-photon mode
  splitting in a semiconductor quantum microcavity},\ }\href
  {https://doi.org/10.1103/physrevlett.69.3314} {\bibfield  {journal} {\bibinfo
   {journal} {Physical Review Letters}\ }\textbf {\bibinfo {volume} {69}},\
  \bibinfo {pages} {3314} (\bibinfo {year} {1992})}\BibitemShut {NoStop}%
\bibitem [{\citenamefont {Kasprzak}\ \emph {et~al.}(2006)\citenamefont
  {Kasprzak}, \citenamefont {Richard}, \citenamefont {Kundermann},
  \citenamefont {Baas}, \citenamefont {Jeambrun}, \citenamefont {Keeling},
  \citenamefont {Marchetti}, \citenamefont {Szyma{\'{n}}ska}, \citenamefont
  {Andr{\'{e}}}, \citenamefont {Staehli}, \citenamefont {Savona}, \citenamefont
  {Littlewood}, \citenamefont {Deveaud},\ and\ \citenamefont
  {Dang}}]{Kasprzak2006}%
  \BibitemOpen
  \bibfield  {author} {\bibinfo {author} {\bibfnamefont {J.}~\bibnamefont
  {Kasprzak}}, \bibinfo {author} {\bibfnamefont {M.}~\bibnamefont {Richard}},
  \bibinfo {author} {\bibfnamefont {S.}~\bibnamefont {Kundermann}}, \bibinfo
  {author} {\bibfnamefont {A.}~\bibnamefont {Baas}}, \bibinfo {author}
  {\bibfnamefont {P.}~\bibnamefont {Jeambrun}}, \bibinfo {author}
  {\bibfnamefont {J.~M.~J.}\ \bibnamefont {Keeling}}, \bibinfo {author}
  {\bibfnamefont {F.~M.}\ \bibnamefont {Marchetti}}, \bibinfo {author}
  {\bibfnamefont {M.~H.}\ \bibnamefont {Szyma{\'{n}}ska}}, \bibinfo {author}
  {\bibfnamefont {R.}~\bibnamefont {Andr{\'{e}}}}, \bibinfo {author}
  {\bibfnamefont {J.~L.}\ \bibnamefont {Staehli}}, \bibinfo {author}
  {\bibfnamefont {V.}~\bibnamefont {Savona}}, \bibinfo {author} {\bibfnamefont
  {P.~B.}\ \bibnamefont {Littlewood}}, \bibinfo {author} {\bibfnamefont
  {B.}~\bibnamefont {Deveaud}},\ and\ \bibinfo {author} {\bibfnamefont {L.~S.}\
  \bibnamefont {Dang}},\ }\bibfield  {title} {\bibinfo {title}
  {Bose{\textendash}{E}instein condensation of exciton polaritons},\ }\href
  {https://doi.org/10.1038/nature05131} {\bibfield  {journal} {\bibinfo
  {journal} {Nature}\ }\textbf {\bibinfo {volume} {443}},\ \bibinfo {pages}
  {409} (\bibinfo {year} {2006})}\BibitemShut {NoStop}%
\bibitem [{\citenamefont {Amo}\ \emph {et~al.}(2009)\citenamefont {Amo},
  \citenamefont {Lefr{\`{e}}re}, \citenamefont {Pigeon}, \citenamefont
  {Adrados}, \citenamefont {Ciuti}, \citenamefont {Carusotto}, \citenamefont
  {Houdr{\'{e}}}, \citenamefont {Giacobino},\ and\ \citenamefont
  {Bramati}}]{Amo2009}%
  \BibitemOpen
  \bibfield  {author} {\bibinfo {author} {\bibfnamefont {A.}~\bibnamefont
  {Amo}}, \bibinfo {author} {\bibfnamefont {J.}~\bibnamefont {Lefr{\`{e}}re}},
  \bibinfo {author} {\bibfnamefont {S.}~\bibnamefont {Pigeon}}, \bibinfo
  {author} {\bibfnamefont {C.}~\bibnamefont {Adrados}}, \bibinfo {author}
  {\bibfnamefont {C.}~\bibnamefont {Ciuti}}, \bibinfo {author} {\bibfnamefont
  {I.}~\bibnamefont {Carusotto}}, \bibinfo {author} {\bibfnamefont
  {R.}~\bibnamefont {Houdr{\'{e}}}}, \bibinfo {author} {\bibfnamefont
  {E.}~\bibnamefont {Giacobino}},\ and\ \bibinfo {author} {\bibfnamefont
  {A.}~\bibnamefont {Bramati}},\ }\bibfield  {title} {\bibinfo {title}
  {Superfluidity of polaritons in semiconductor microcavities},\ }\href
  {https://doi.org/10.1038/nphys1364} {\bibfield  {journal} {\bibinfo
  {journal} {Nature Physics}\ }\textbf {\bibinfo {volume} {5}},\ \bibinfo
  {pages} {805} (\bibinfo {year} {2009})}\BibitemShut {NoStop}%
\bibitem [{\citenamefont {Amo}\ \emph {et~al.}(2011)\citenamefont {Amo},
  \citenamefont {Pigeon}, \citenamefont {Sanvitto}, \citenamefont {Sala},
  \citenamefont {Hivet}, \citenamefont {Carusotto}, \citenamefont {Pisanello},
  \citenamefont {Lem{\'{e}}nager}, \citenamefont {Houdr{\'{e}}}, \citenamefont
  {Giacobino}, \citenamefont {Ciuti},\ and\ \citenamefont {Bramati}}]{Amo2011}%
  \BibitemOpen
  \bibfield  {author} {\bibinfo {author} {\bibfnamefont {A.}~\bibnamefont
  {Amo}}, \bibinfo {author} {\bibfnamefont {S.}~\bibnamefont {Pigeon}},
  \bibinfo {author} {\bibfnamefont {D.}~\bibnamefont {Sanvitto}}, \bibinfo
  {author} {\bibfnamefont {V.~G.}\ \bibnamefont {Sala}}, \bibinfo {author}
  {\bibfnamefont {R.}~\bibnamefont {Hivet}}, \bibinfo {author} {\bibfnamefont
  {I.}~\bibnamefont {Carusotto}}, \bibinfo {author} {\bibfnamefont
  {F.}~\bibnamefont {Pisanello}}, \bibinfo {author} {\bibfnamefont
  {G.}~\bibnamefont {Lem{\'{e}}nager}}, \bibinfo {author} {\bibfnamefont
  {R.}~\bibnamefont {Houdr{\'{e}}}}, \bibinfo {author} {\bibfnamefont
  {E.}~\bibnamefont {Giacobino}}, \bibinfo {author} {\bibfnamefont
  {C.}~\bibnamefont {Ciuti}},\ and\ \bibinfo {author} {\bibfnamefont
  {A.}~\bibnamefont {Bramati}},\ }\bibfield  {title} {\bibinfo {title}
  {Polariton superfluids reveal quantum hydrodynamic solitons},\ }\href
  {https://doi.org/10.1126/science.1202307} {\bibfield  {journal} {\bibinfo
  {journal} {Science}\ }\textbf {\bibinfo {volume} {332}},\ \bibinfo {pages}
  {1167} (\bibinfo {year} {2011})}\BibitemShut {NoStop}%
\bibitem [{\citenamefont {Barland}\ \emph {et~al.}(2012)\citenamefont
  {Barland}, \citenamefont {Giudici}, \citenamefont {Tissoni}, \citenamefont
  {Tredicce}, \citenamefont {Brambilla}, \citenamefont {Lugiato}, \citenamefont
  {Prati}, \citenamefont {Barbay}, \citenamefont {Kuszelewicz}, \citenamefont
  {Ackemann}, \citenamefont {Firth},\ and\ \citenamefont {Oppo}}]{Barland2012}%
  \BibitemOpen
  \bibfield  {author} {\bibinfo {author} {\bibfnamefont {S.}~\bibnamefont
  {Barland}}, \bibinfo {author} {\bibfnamefont {M.}~\bibnamefont {Giudici}},
  \bibinfo {author} {\bibfnamefont {G.}~\bibnamefont {Tissoni}}, \bibinfo
  {author} {\bibfnamefont {J.~R.}\ \bibnamefont {Tredicce}}, \bibinfo {author}
  {\bibfnamefont {M.}~\bibnamefont {Brambilla}}, \bibinfo {author}
  {\bibfnamefont {L.}~\bibnamefont {Lugiato}}, \bibinfo {author} {\bibfnamefont
  {F.}~\bibnamefont {Prati}}, \bibinfo {author} {\bibfnamefont
  {S.}~\bibnamefont {Barbay}}, \bibinfo {author} {\bibfnamefont
  {R.}~\bibnamefont {Kuszelewicz}}, \bibinfo {author} {\bibfnamefont
  {T.}~\bibnamefont {Ackemann}}, \bibinfo {author} {\bibfnamefont {W.~J.}\
  \bibnamefont {Firth}},\ and\ \bibinfo {author} {\bibfnamefont {G.-L.}\
  \bibnamefont {Oppo}},\ }\bibfield  {title} {\bibinfo {title} {Solitons in
  semiconductor microcavities},\ }\href
  {https://doi.org/10.1038/nphoton.2012.50} {\bibfield  {journal} {\bibinfo
  {journal} {Nature Photonics}\ }\textbf {\bibinfo {volume} {6}},\ \bibinfo
  {pages} {204} (\bibinfo {year} {2012})}\BibitemShut {NoStop}%
\bibitem [{\citenamefont {Lagoudakis}\ \emph {et~al.}(2008)\citenamefont
  {Lagoudakis}, \citenamefont {Wouters}, \citenamefont {Richard}, \citenamefont
  {Baas}, \citenamefont {Carusotto}, \citenamefont {Andr{\'{e}}}, \citenamefont
  {Dang},\ and\ \citenamefont {Deveaud-Pl{\'{e}}dran}}]{Lagoudakis2008}%
  \BibitemOpen
  \bibfield  {author} {\bibinfo {author} {\bibfnamefont {K.~G.}\ \bibnamefont
  {Lagoudakis}}, \bibinfo {author} {\bibfnamefont {M.}~\bibnamefont {Wouters}},
  \bibinfo {author} {\bibfnamefont {M.}~\bibnamefont {Richard}}, \bibinfo
  {author} {\bibfnamefont {A.}~\bibnamefont {Baas}}, \bibinfo {author}
  {\bibfnamefont {I.}~\bibnamefont {Carusotto}}, \bibinfo {author}
  {\bibfnamefont {R.}~\bibnamefont {Andr{\'{e}}}}, \bibinfo {author}
  {\bibfnamefont {L.~S.}\ \bibnamefont {Dang}},\ and\ \bibinfo {author}
  {\bibfnamefont {B.}~\bibnamefont {Deveaud-Pl{\'{e}}dran}},\ }\bibfield
  {title} {\bibinfo {title} {Quantized vortices in an
  exciton{\textendash}polariton condensate},\ }\href
  {https://doi.org/10.1038/nphys1051} {\bibfield  {journal} {\bibinfo
  {journal} {Nature Physics}\ }\textbf {\bibinfo {volume} {4}},\ \bibinfo
  {pages} {706} (\bibinfo {year} {2008})}\BibitemShut {NoStop}%
\bibitem [{\citenamefont {Kardar}\ \emph {et~al.}(1986)\citenamefont {Kardar},
  \citenamefont {Parisi},\ and\ \citenamefont {Zhang}}]{Kardar1986}%
  \BibitemOpen
  \bibfield  {author} {\bibinfo {author} {\bibfnamefont {M.}~\bibnamefont
  {Kardar}}, \bibinfo {author} {\bibfnamefont {G.}~\bibnamefont {Parisi}},\
  and\ \bibinfo {author} {\bibfnamefont {Y.-C.}\ \bibnamefont {Zhang}},\
  }\bibfield  {title} {\bibinfo {title} {Dynamic scaling of growing
  interfaces},\ }\href {https://doi.org/10.1103/physrevlett.56.889} {\bibfield
  {journal} {\bibinfo  {journal} {Physical Review Letters}\ }\textbf {\bibinfo
  {volume} {56}},\ \bibinfo {pages} {889} (\bibinfo {year} {1986})}\BibitemShut
  {NoStop}%
\bibitem [{\citenamefont {Takeuchi}(2018)}]{Takeuchi2018}%
  \BibitemOpen
  \bibfield  {author} {\bibinfo {author} {\bibfnamefont {K.~A.}\ \bibnamefont
  {Takeuchi}},\ }\bibfield  {title} {\bibinfo {title} {An appetizer to modern
  developments on the {K}ardar{\textendash}{P}arisi{\textendash}{Z}hang
  universality class},\ }\href {https://doi.org/10.1016/j.physa.2018.03.009}
  {\bibfield  {journal} {\bibinfo  {journal} {Physica A: Statistical Mechanics
  and its Applications}\ }\textbf {\bibinfo {volume} {504}},\ \bibinfo {pages}
  {77} (\bibinfo {year} {2018})}\BibitemShut {NoStop}%
\bibitem [{\citenamefont {Fontaine}\ \emph {et~al.}(2022)\citenamefont
  {Fontaine}, \citenamefont {Squizzato}, \citenamefont {Baboux}, \citenamefont
  {Amelio}, \citenamefont {Lema{\^{\i}}tre}, \citenamefont {Morassi},
  \citenamefont {Sagnes}, \citenamefont {Gratiet}, \citenamefont {Harouri},
  \citenamefont {Wouters}, \citenamefont {Carusotto}, \citenamefont {Amo},
  \citenamefont {Richard}, \citenamefont {Minguzzi}, \citenamefont {Canet},
  \citenamefont {Ravets},\ and\ \citenamefont {Bloch}}]{Fontaine2022}%
  \BibitemOpen
  \bibfield  {author} {\bibinfo {author} {\bibfnamefont {Q.}~\bibnamefont
  {Fontaine}}, \bibinfo {author} {\bibfnamefont {D.}~\bibnamefont {Squizzato}},
  \bibinfo {author} {\bibfnamefont {F.}~\bibnamefont {Baboux}}, \bibinfo
  {author} {\bibfnamefont {I.}~\bibnamefont {Amelio}}, \bibinfo {author}
  {\bibfnamefont {A.}~\bibnamefont {Lema{\^{\i}}tre}}, \bibinfo {author}
  {\bibfnamefont {M.}~\bibnamefont {Morassi}}, \bibinfo {author} {\bibfnamefont
  {I.}~\bibnamefont {Sagnes}}, \bibinfo {author} {\bibfnamefont {L.~L.}\
  \bibnamefont {Gratiet}}, \bibinfo {author} {\bibfnamefont {A.}~\bibnamefont
  {Harouri}}, \bibinfo {author} {\bibfnamefont {M.}~\bibnamefont {Wouters}},
  \bibinfo {author} {\bibfnamefont {I.}~\bibnamefont {Carusotto}}, \bibinfo
  {author} {\bibfnamefont {A.}~\bibnamefont {Amo}}, \bibinfo {author}
  {\bibfnamefont {M.}~\bibnamefont {Richard}}, \bibinfo {author} {\bibfnamefont
  {A.}~\bibnamefont {Minguzzi}}, \bibinfo {author} {\bibfnamefont
  {L.}~\bibnamefont {Canet}}, \bibinfo {author} {\bibfnamefont
  {S.}~\bibnamefont {Ravets}},\ and\ \bibinfo {author} {\bibfnamefont
  {J.}~\bibnamefont {Bloch}},\ }\bibfield  {title} {\bibinfo {title}
  {{K}ardar{\textendash}{P}arisi{\textendash}{Z}hang universality in a
  one-dimensional polariton condensate},\ }\href
  {https://doi.org/10.1038/s41586-022-05001-8} {\bibfield  {journal} {\bibinfo
  {journal} {Nature}\ }\textbf {\bibinfo {volume} {608}},\ \bibinfo {pages}
  {687} (\bibinfo {year} {2022})}\BibitemShut {NoStop}%
\bibitem [{\citenamefont {He}\ \emph {et~al.}(2015)\citenamefont {He},
  \citenamefont {Sieberer}, \citenamefont {Altman},\ and\ \citenamefont
  {Diehl}}]{He2015}%
  \BibitemOpen
  \bibfield  {author} {\bibinfo {author} {\bibfnamefont {L.}~\bibnamefont
  {He}}, \bibinfo {author} {\bibfnamefont {L.~M.}\ \bibnamefont {Sieberer}},
  \bibinfo {author} {\bibfnamefont {E.}~\bibnamefont {Altman}},\ and\ \bibinfo
  {author} {\bibfnamefont {S.}~\bibnamefont {Diehl}},\ }\bibfield  {title}
  {\bibinfo {title} {Scaling properties of one-dimensional driven-dissipative
  condensates},\ }\bibfield  {journal} {\bibinfo  {journal} {Physical Review
  B}\ }\textbf {\bibinfo {volume} {92}},\ \href
  {https://doi.org/10.1103/physrevb.92.155307} {10.1103/physrevb.92.155307}
  (\bibinfo {year} {2015})\BibitemShut {NoStop}%
\bibitem [{\citenamefont {Ji}\ \emph {et~al.}(2015)\citenamefont {Ji},
  \citenamefont {Gladilin},\ and\ \citenamefont {Wouters}}]{Ji2015}%
  \BibitemOpen
  \bibfield  {author} {\bibinfo {author} {\bibfnamefont {K.}~\bibnamefont
  {Ji}}, \bibinfo {author} {\bibfnamefont {V.~N.}\ \bibnamefont {Gladilin}},\
  and\ \bibinfo {author} {\bibfnamefont {M.}~\bibnamefont {Wouters}},\
  }\bibfield  {title} {\bibinfo {title} {Temporal coherence of one-dimensional
  nonequilibrium quantum fluids},\ }\bibfield  {journal} {\bibinfo  {journal}
  {Physical Review B}\ }\textbf {\bibinfo {volume} {91}},\ \href
  {https://doi.org/10.1103/physrevb.91.045301} {10.1103/physrevb.91.045301}
  (\bibinfo {year} {2015})\BibitemShut {NoStop}%
\bibitem [{\citenamefont {Squizzato}\ \emph {et~al.}(2018)\citenamefont
  {Squizzato}, \citenamefont {Canet},\ and\ \citenamefont
  {Minguzzi}}]{Squizzato2018}%
  \BibitemOpen
  \bibfield  {author} {\bibinfo {author} {\bibfnamefont {D.}~\bibnamefont
  {Squizzato}}, \bibinfo {author} {\bibfnamefont {L.}~\bibnamefont {Canet}},\
  and\ \bibinfo {author} {\bibfnamefont {A.}~\bibnamefont {Minguzzi}},\
  }\bibfield  {title} {\bibinfo {title} {Kardar-{P}arisi-{Z}hang universality
  in the phase distributions of one-dimensional exciton-polaritons},\
  }\bibfield  {journal} {\bibinfo  {journal} {Physical Review B}\ }\textbf
  {\bibinfo {volume} {97}},\ \href {https://doi.org/10.1103/physrevb.97.195453}
  {10.1103/physrevb.97.195453} (\bibinfo {year} {2018})\BibitemShut {NoStop}%
\bibitem [{\citenamefont {Deligiannis}\ \emph {et~al.}(2021)\citenamefont
  {Deligiannis}, \citenamefont {Squizzato}, \citenamefont {Minguzzi},\ and\
  \citenamefont {Canet}}]{Deligiannis2021}%
  \BibitemOpen
  \bibfield  {author} {\bibinfo {author} {\bibfnamefont {K.}~\bibnamefont
  {Deligiannis}}, \bibinfo {author} {\bibfnamefont {D.}~\bibnamefont
  {Squizzato}}, \bibinfo {author} {\bibfnamefont {A.}~\bibnamefont
  {Minguzzi}},\ and\ \bibinfo {author} {\bibfnamefont {L.}~\bibnamefont
  {Canet}},\ }\bibfield  {title} {\bibinfo {title} {Accessing
  {K}ardar-{P}arisi-{Z}hang universality sub-classes with exciton polaritons},\
  }\href {https://doi.org/10.1209/0295-5075/132/67004} {\bibfield  {journal}
  {\bibinfo  {journal} {{EPL} (Europhysics Letters)}\ }\textbf {\bibinfo
  {volume} {132}},\ \bibinfo {pages} {67004} (\bibinfo {year}
  {2021})}\BibitemShut {NoStop}%
\bibitem [{\citenamefont {Bobrovska}\ \emph {et~al.}(2014)\citenamefont
  {Bobrovska}, \citenamefont {Ostrovskaya},\ and\ \citenamefont
  {Matuszewski}}]{Bobrovska2014}%
  \BibitemOpen
  \bibfield  {author} {\bibinfo {author} {\bibfnamefont {N.}~\bibnamefont
  {Bobrovska}}, \bibinfo {author} {\bibfnamefont {E.~A.}\ \bibnamefont
  {Ostrovskaya}},\ and\ \bibinfo {author} {\bibfnamefont {M.}~\bibnamefont
  {Matuszewski}},\ }\bibfield  {title} {\bibinfo {title} {Stability and spatial
  coherence of nonresonantly pumped exciton-polariton condensates},\ }\bibfield
   {journal} {\bibinfo  {journal} {Physical Review B}\ }\textbf {\bibinfo
  {volume} {90}},\ \href {https://doi.org/10.1103/physrevb.90.205304}
  {10.1103/physrevb.90.205304} (\bibinfo {year} {2014})\BibitemShut {NoStop}%
\bibitem [{\citenamefont {Aranson}\ and\ \citenamefont
  {Kramer}(2002)}]{Aranson2002}%
  \BibitemOpen
  \bibfield  {author} {\bibinfo {author} {\bibfnamefont {I.~S.}\ \bibnamefont
  {Aranson}}\ and\ \bibinfo {author} {\bibfnamefont {L.}~\bibnamefont
  {Kramer}},\ }\bibfield  {title} {\bibinfo {title} {The world of the complex
  {G}inzburg-{L}andau equation},\ }\href
  {https://doi.org/10.1103/revmodphys.74.99} {\bibfield  {journal} {\bibinfo
  {journal} {Reviews of Modern Physics}\ }\textbf {\bibinfo {volume} {74}},\
  \bibinfo {pages} {99} (\bibinfo {year} {2002})}\BibitemShut {NoStop}%
\bibitem [{\citenamefont {Chat{\'{e}}}(1994)}]{Chate1994}%
  \BibitemOpen
  \bibfield  {author} {\bibinfo {author} {\bibfnamefont {H.}~\bibnamefont
  {Chat{\'{e}}}},\ }\bibfield  {title} {\bibinfo {title} {Spatiotemporal
  intermittency regimes of the one-dimensional complex {G}inzburg-{L}andau
  equation},\ }\href {https://doi.org/10.1088/0951-7715/7/1/007} {\bibfield
  {journal} {\bibinfo  {journal} {Nonlinearity}\ }\textbf {\bibinfo {volume}
  {7}},\ \bibinfo {pages} {185} (\bibinfo {year} {1994})}\BibitemShut {NoStop}%
\bibitem [{\citenamefont {Chat{\'{e}}}(1995)}]{Chat1995}%
  \BibitemOpen
  \bibfield  {author} {\bibinfo {author} {\bibfnamefont {H.}~\bibnamefont
  {Chat{\'{e}}}},\ }\bibfield  {title} {\bibinfo {title} {On the analysis of
  spatiotemporally chaotic data},\ }\href
  {https://doi.org/10.1016/0167-2789(95)00104-c} {\bibfield  {journal}
  {\bibinfo  {journal} {Physica D: Nonlinear Phenomena}\ }\textbf {\bibinfo
  {volume} {86}},\ \bibinfo {pages} {238} (\bibinfo {year} {1995})}\BibitemShut
  {NoStop}%
\bibitem [{\citenamefont {Grinstein}\ \emph {et~al.}(1996)\citenamefont
  {Grinstein}, \citenamefont {Jayaprakash},\ and\ \citenamefont
  {Pandit}}]{Grinstein1996}%
  \BibitemOpen
  \bibfield  {author} {\bibinfo {author} {\bibfnamefont {G.}~\bibnamefont
  {Grinstein}}, \bibinfo {author} {\bibfnamefont {C.}~\bibnamefont
  {Jayaprakash}},\ and\ \bibinfo {author} {\bibfnamefont {R.}~\bibnamefont
  {Pandit}},\ }\bibfield  {title} {\bibinfo {title} {Conjectures about phase
  turbulence in the complex {G}inzburg{\textendash}{L}andau equation},\ }\href
  {https://doi.org/10.1016/0167-2789(95)00036-4} {\bibfield  {journal}
  {\bibinfo  {journal} {Physica D: Nonlinear Phenomena}\ }\textbf {\bibinfo
  {volume} {90}},\ \bibinfo {pages} {96} (\bibinfo {year} {1996})}\BibitemShut
  {NoStop}%
\bibitem [{\citenamefont {Pernet}\ \emph {et~al.}(2022)\citenamefont {Pernet},
  \citenamefont {St-Jean}, \citenamefont {Solnyshkov}, \citenamefont
  {Malpuech}, \citenamefont {Zambon}, \citenamefont {Fontaine}, \citenamefont
  {Real}, \citenamefont {Jamadi}, \citenamefont {Lema{\^{\i}}tre},
  \citenamefont {Morassi}, \citenamefont {Gratiet}, \citenamefont {Baptiste},
  \citenamefont {Harouri}, \citenamefont {Sagnes}, \citenamefont {Amo},
  \citenamefont {Ravets},\ and\ \citenamefont {Bloch}}]{Pernet2022}%
  \BibitemOpen
  \bibfield  {author} {\bibinfo {author} {\bibfnamefont {N.}~\bibnamefont
  {Pernet}}, \bibinfo {author} {\bibfnamefont {P.}~\bibnamefont {St-Jean}},
  \bibinfo {author} {\bibfnamefont {D.~D.}\ \bibnamefont {Solnyshkov}},
  \bibinfo {author} {\bibfnamefont {G.}~\bibnamefont {Malpuech}}, \bibinfo
  {author} {\bibfnamefont {N.~C.}\ \bibnamefont {Zambon}}, \bibinfo {author}
  {\bibfnamefont {Q.}~\bibnamefont {Fontaine}}, \bibinfo {author}
  {\bibfnamefont {B.}~\bibnamefont {Real}}, \bibinfo {author} {\bibfnamefont
  {O.}~\bibnamefont {Jamadi}}, \bibinfo {author} {\bibfnamefont
  {A.}~\bibnamefont {Lema{\^{\i}}tre}}, \bibinfo {author} {\bibfnamefont
  {M.}~\bibnamefont {Morassi}}, \bibinfo {author} {\bibfnamefont {L.~L.}\
  \bibnamefont {Gratiet}}, \bibinfo {author} {\bibfnamefont {T.}~\bibnamefont
  {Baptiste}}, \bibinfo {author} {\bibfnamefont {A.}~\bibnamefont {Harouri}},
  \bibinfo {author} {\bibfnamefont {I.}~\bibnamefont {Sagnes}}, \bibinfo
  {author} {\bibfnamefont {A.}~\bibnamefont {Amo}}, \bibinfo {author}
  {\bibfnamefont {S.}~\bibnamefont {Ravets}},\ and\ \bibinfo {author}
  {\bibfnamefont {J.}~\bibnamefont {Bloch}},\ }\bibfield  {title} {\bibinfo
  {title} {Gap solitons in a one-dimensional driven-dissipative topological
  lattice},\ }\href {https://doi.org/10.1038/s41567-022-01599-8} {\bibfield
  {journal} {\bibinfo  {journal} {Nature Physics}\ }\textbf {\bibinfo {volume}
  {18}},\ \bibinfo {pages} {678} (\bibinfo {year} {2022})}\BibitemShut
  {NoStop}%
\bibitem [{\citenamefont {Bobrovska}\ and\ \citenamefont
  {Matuszewski}(2015)}]{Bobrovska2015}%
  \BibitemOpen
  \bibfield  {author} {\bibinfo {author} {\bibfnamefont {N.}~\bibnamefont
  {Bobrovska}}\ and\ \bibinfo {author} {\bibfnamefont {M.}~\bibnamefont
  {Matuszewski}},\ }\bibfield  {title} {\bibinfo {title} {Adiabatic
  approximation and fluctuations in exciton-polariton condensates},\ }\bibfield
   {journal} {\bibinfo  {journal} {Physical Review B}\ }\textbf {\bibinfo
  {volume} {92}},\ \href {https://doi.org/10.1103/physrevb.92.035311}
  {10.1103/physrevb.92.035311} (\bibinfo {year} {2015})\BibitemShut {NoStop}%
\bibitem [{\citenamefont {Bobrovska}\ \emph {et~al.}(2017)\citenamefont
  {Bobrovska}, \citenamefont {Matuszewski}, \citenamefont {Daskalakis},
  \citenamefont {Maier},\ and\ \citenamefont
  {K{\'{e}}na-Cohen}}]{Bobrovska2017}%
  \BibitemOpen
  \bibfield  {author} {\bibinfo {author} {\bibfnamefont {N.}~\bibnamefont
  {Bobrovska}}, \bibinfo {author} {\bibfnamefont {M.}~\bibnamefont
  {Matuszewski}}, \bibinfo {author} {\bibfnamefont {K.~S.}\ \bibnamefont
  {Daskalakis}}, \bibinfo {author} {\bibfnamefont {S.~A.}\ \bibnamefont
  {Maier}},\ and\ \bibinfo {author} {\bibfnamefont {S.}~\bibnamefont
  {K{\'{e}}na-Cohen}},\ }\bibfield  {title} {\bibinfo {title} {Dynamical
  instability of a nonequilibrium exciton-polariton condensate},\ }\href
  {https://doi.org/10.1021/acsphotonics.7b00283} {\bibfield  {journal}
  {\bibinfo  {journal} {{ACS} Photonics}\ }\textbf {\bibinfo {volume} {5}},\
  \bibinfo {pages} {111} (\bibinfo {year} {2017})}\BibitemShut {NoStop}%
\bibitem [{\citenamefont {Baboux}\ \emph {et~al.}(2018)\citenamefont {Baboux},
  \citenamefont {Bernardis}, \citenamefont {Goblot}, \citenamefont {Gladilin},
  \citenamefont {Gomez}, \citenamefont {Galopin}, \citenamefont {Gratiet},
  \citenamefont {Lema{\^{\i}}tre}, \citenamefont {Sagnes}, \citenamefont
  {Carusotto}, \citenamefont {Wouters}, \citenamefont {Amo},\ and\
  \citenamefont {Bloch}}]{Baboux2018}%
  \BibitemOpen
  \bibfield  {author} {\bibinfo {author} {\bibfnamefont {F.}~\bibnamefont
  {Baboux}}, \bibinfo {author} {\bibfnamefont {D.~D.}\ \bibnamefont
  {Bernardis}}, \bibinfo {author} {\bibfnamefont {V.}~\bibnamefont {Goblot}},
  \bibinfo {author} {\bibfnamefont {V.~N.}\ \bibnamefont {Gladilin}}, \bibinfo
  {author} {\bibfnamefont {C.}~\bibnamefont {Gomez}}, \bibinfo {author}
  {\bibfnamefont {E.}~\bibnamefont {Galopin}}, \bibinfo {author} {\bibfnamefont
  {L.~L.}\ \bibnamefont {Gratiet}}, \bibinfo {author} {\bibfnamefont
  {A.}~\bibnamefont {Lema{\^{\i}}tre}}, \bibinfo {author} {\bibfnamefont
  {I.}~\bibnamefont {Sagnes}}, \bibinfo {author} {\bibfnamefont
  {I.}~\bibnamefont {Carusotto}}, \bibinfo {author} {\bibfnamefont
  {M.}~\bibnamefont {Wouters}}, \bibinfo {author} {\bibfnamefont
  {A.}~\bibnamefont {Amo}},\ and\ \bibinfo {author} {\bibfnamefont
  {J.}~\bibnamefont {Bloch}},\ }\bibfield  {title} {\bibinfo {title} {Unstable
  and stable regimes of polariton condensation},\ }\href
  {https://doi.org/10.1364/optica.5.001163} {\bibfield  {journal} {\bibinfo
  {journal} {Optica}\ }\textbf {\bibinfo {volume} {5}},\ \bibinfo {pages}
  {1163} (\bibinfo {year} {2018})}\BibitemShut {NoStop}%
\bibitem [{\citenamefont {Bobrovska}\ \emph {et~al.}(2019)\citenamefont
  {Bobrovska}, \citenamefont {Opala}, \citenamefont {Mi{\k{e}}tki},
  \citenamefont {Kulczykowski}, \citenamefont {Szymczak}, \citenamefont
  {Wouters},\ and\ \citenamefont {Matuszewski}}]{Bobrovska2019}%
  \BibitemOpen
  \bibfield  {author} {\bibinfo {author} {\bibfnamefont {N.}~\bibnamefont
  {Bobrovska}}, \bibinfo {author} {\bibfnamefont {A.}~\bibnamefont {Opala}},
  \bibinfo {author} {\bibfnamefont {P.}~\bibnamefont {Mi{\k{e}}tki}}, \bibinfo
  {author} {\bibfnamefont {M.}~\bibnamefont {Kulczykowski}}, \bibinfo {author}
  {\bibfnamefont {P.}~\bibnamefont {Szymczak}}, \bibinfo {author}
  {\bibfnamefont {M.}~\bibnamefont {Wouters}},\ and\ \bibinfo {author}
  {\bibfnamefont {M.}~\bibnamefont {Matuszewski}},\ }\bibfield  {title}
  {\bibinfo {title} {Critical dynamics and tree-like spatiotemporal patterns in
  exciton-polariton condensates},\ }\bibfield  {journal} {\bibinfo  {journal}
  {Physical Review B}\ }\textbf {\bibinfo {volume} {99}},\ \href
  {https://doi.org/10.1103/physrevb.99.205301} {10.1103/physrevb.99.205301}
  (\bibinfo {year} {2019})\BibitemShut {NoStop}%
\bibitem [{\citenamefont {Smirnov}\ \emph {et~al.}(2014)\citenamefont
  {Smirnov}, \citenamefont {Smirnova}, \citenamefont {Ostrovskaya},\ and\
  \citenamefont {Kivshar}}]{Smirnov2014}%
  \BibitemOpen
  \bibfield  {author} {\bibinfo {author} {\bibfnamefont {L.~A.}\ \bibnamefont
  {Smirnov}}, \bibinfo {author} {\bibfnamefont {D.~A.}\ \bibnamefont
  {Smirnova}}, \bibinfo {author} {\bibfnamefont {E.~A.}\ \bibnamefont
  {Ostrovskaya}},\ and\ \bibinfo {author} {\bibfnamefont {Y.~S.}\ \bibnamefont
  {Kivshar}},\ }\bibfield  {title} {\bibinfo {title} {Dynamics and stability of
  dark solitons in exciton-polariton condensates},\ }\bibfield  {journal}
  {\bibinfo  {journal} {Physical Review B}\ }\textbf {\bibinfo {volume} {89}},\
  \href {https://doi.org/10.1103/physrevb.89.235310}
  {10.1103/physrevb.89.235310} (\bibinfo {year} {2014})\BibitemShut {NoStop}%
\bibitem [{\citenamefont {Opala}\ \emph {et~al.}(2018)\citenamefont {Opala},
  \citenamefont {Pieczarka}, \citenamefont {Bobrovska},\ and\ \citenamefont
  {Matuszewski}}]{Opala2018}%
  \BibitemOpen
  \bibfield  {author} {\bibinfo {author} {\bibfnamefont {A.}~\bibnamefont
  {Opala}}, \bibinfo {author} {\bibfnamefont {M.}~\bibnamefont {Pieczarka}},
  \bibinfo {author} {\bibfnamefont {N.}~\bibnamefont {Bobrovska}},\ and\
  \bibinfo {author} {\bibfnamefont {M.}~\bibnamefont {Matuszewski}},\
  }\bibfield  {title} {\bibinfo {title} {Dynamics of defect-induced dark
  solitons in an exciton-polariton condensate},\ }\bibfield  {journal}
  {\bibinfo  {journal} {Physical Review B}\ }\textbf {\bibinfo {volume} {97}},\
  \href {https://doi.org/10.1103/physrevb.97.155304}
  {10.1103/physrevb.97.155304} (\bibinfo {year} {2018})\BibitemShut {NoStop}%
\bibitem [{\citenamefont {He}\ \emph {et~al.}(2017)\citenamefont {He},
  \citenamefont {Sieberer},\ and\ \citenamefont {Diehl}}]{He2017}%
  \BibitemOpen
  \bibfield  {author} {\bibinfo {author} {\bibfnamefont {L.}~\bibnamefont
  {He}}, \bibinfo {author} {\bibfnamefont {L.~M.}\ \bibnamefont {Sieberer}},\
  and\ \bibinfo {author} {\bibfnamefont {S.}~\bibnamefont {Diehl}},\ }\bibfield
   {title} {\bibinfo {title} {Space-time vortex driven crossover and vortex
  turbulence phase transition in one-dimensional driven open condensates},\
  }\bibfield  {journal} {\bibinfo  {journal} {Physical Review Letters}\
  }\textbf {\bibinfo {volume} {118}},\ \href
  {https://doi.org/10.1103/physrevlett.118.085301}
  {10.1103/physrevlett.118.085301} (\bibinfo {year} {2017})\BibitemShut
  {NoStop}%
\bibitem [{\citenamefont {Wouters}\ and\ \citenamefont
  {Savona}(2009)}]{Wouters2009}%
  \BibitemOpen
  \bibfield  {author} {\bibinfo {author} {\bibfnamefont {M.}~\bibnamefont
  {Wouters}}\ and\ \bibinfo {author} {\bibfnamefont {V.}~\bibnamefont
  {Savona}},\ }\bibfield  {title} {\bibinfo {title} {Stochastic classical field
  model for polariton condensates},\ }\bibfield  {journal} {\bibinfo  {journal}
  {Physical Review B}\ }\textbf {\bibinfo {volume} {79}},\ \href
  {https://doi.org/10.1103/physrevb.79.165302} {10.1103/physrevb.79.165302}
  (\bibinfo {year} {2009})\BibitemShut {NoStop}%
\bibitem [{\citenamefont {Pr\"{a}hofer}\ and\ \citenamefont
  {Spohn}(2004)}]{Prhofer2004}%
  \BibitemOpen
  \bibfield  {author} {\bibinfo {author} {\bibfnamefont {M.}~\bibnamefont
  {Pr\"{a}hofer}}\ and\ \bibinfo {author} {\bibfnamefont {H.}~\bibnamefont
  {Spohn}},\ }\bibfield  {title} {\bibinfo {title} {Exact scaling functions for
  one-dimensional stationary {KPZ} growth},\ }\href
  {https://doi.org/10.1023/b:joss.0000019810.21828.fc} {\bibfield  {journal}
  {\bibinfo  {journal} {Journal of Statistical Physics}\ }\textbf {\bibinfo
  {volume} {115}},\ \bibinfo {pages} {255} (\bibinfo {year}
  {2004})}\BibitemShut {NoStop}%
\bibitem [{\citenamefont {Leggett}(2004)}]{Leggett2004}%
  \BibitemOpen
  \bibfield  {author} {\bibinfo {author} {\bibfnamefont {A.}~\bibnamefont
  {Leggett}},\ }\href {https://doi.org/10.1063/1.1825272} {\emph {\bibinfo
  {title} {{B}ose{\textendash}{E}instein Condensation}}},\ edited by\ \bibinfo
  {editor} {\bibfnamefont {L.}~\bibnamefont {Pitaevskii}}\ and\ \bibinfo
  {editor} {\bibfnamefont {N.~Y.}\ \bibnamefont {Sandro~Stringari},
  \bibfnamefont {Oxford U.~Press}},\ Vol.~\bibinfo {volume} {57}\ (\bibinfo
  {publisher} {{AIP} Publishing},\ \bibinfo {year} {2004})\BibitemShut
  {NoStop}%
\bibitem [{\citenamefont {Cartes}\ \emph {et~al.}(2022)\citenamefont {Cartes},
  \citenamefont {Tirapegui}, \citenamefont {Pandit},\ and\ \citenamefont
  {Brachet}}]{Brachet2022}%
  \BibitemOpen
  \bibfield  {author} {\bibinfo {author} {\bibfnamefont {C.}~\bibnamefont
  {Cartes}}, \bibinfo {author} {\bibfnamefont {E.}~\bibnamefont {Tirapegui}},
  \bibinfo {author} {\bibfnamefont {R.}~\bibnamefont {Pandit}},\ and\ \bibinfo
  {author} {\bibfnamefont {M.}~\bibnamefont {Brachet}},\ }\bibfield  {title}
  {\bibinfo {title} {The {G}alerkin-truncated {B}urgers equation: crossover
  from inviscid-thermalized to {K}ardar–{P}arisi–{Z}hang scaling},\ }\href
  {https://doi.org/10.1098/rsta.2021.0090} {\bibfield  {journal} {\bibinfo
  {journal} {Philosophical Transactions of the Royal Society A: Mathematical,
  Physical and Engineering Sciences}\ }\textbf {\bibinfo {volume} {380}},\
  \bibinfo {pages} {20210090} (\bibinfo {year} {2022})}\BibitemShut {NoStop}%
\bibitem [{\citenamefont {Rodr\'{\i}guez-Fern\'andez}\ \emph
  {et~al.}(2022)\citenamefont {Rodr\'{\i}guez-Fern\'andez}, \citenamefont
  {Santalla}, \citenamefont {Castro},\ and\ \citenamefont
  {Cuerno}}]{Cuerno2022}%
  \BibitemOpen
  \bibfield  {author} {\bibinfo {author} {\bibfnamefont {E.}~\bibnamefont
  {Rodr\'{\i}guez-Fern\'andez}}, \bibinfo {author} {\bibfnamefont {S.~N.}\
  \bibnamefont {Santalla}}, \bibinfo {author} {\bibfnamefont {M.}~\bibnamefont
  {Castro}},\ and\ \bibinfo {author} {\bibfnamefont {R.}~\bibnamefont
  {Cuerno}},\ }\bibfield  {title} {\bibinfo {title} {Anomalous ballistic
  scaling in the tensionless or inviscid {K}ardar-{P}arisi-{Z}hang equation},\
  }\href {https://doi.org/10.1103/PhysRevE.106.024802} {\bibfield  {journal}
  {\bibinfo  {journal} {Phys. Rev. E}\ }\textbf {\bibinfo {volume} {106}},\
  \bibinfo {pages} {024802} (\bibinfo {year} {2022})}\BibitemShut {NoStop}%
\bibitem [{\citenamefont {Fontaine}\ \emph {et~al.}(2023)\citenamefont
  {Fontaine}, \citenamefont {Vercesi}, \citenamefont {Brachet},\ and\
  \citenamefont {Canet}}]{Fontaine2023}%
  \BibitemOpen
  \bibfield  {author} {\bibinfo {author} {\bibfnamefont {C.}~\bibnamefont
  {Fontaine}}, \bibinfo {author} {\bibfnamefont {F.}~\bibnamefont {Vercesi}},
  \bibinfo {author} {\bibfnamefont {M.}~\bibnamefont {Brachet}},\ and\ \bibinfo
  {author} {\bibfnamefont {L.}~\bibnamefont {Canet}},\ }\bibfield  {title}
  {\bibinfo {title} {The unpredicted scaling of the one-dimensional
  {K}ardar-{P}arisi-{Z}hang equation},\ }\href@noop {} {\bibfield  {journal}
  {\bibinfo  {journal} {arXiv:2305.09358}\ } (\bibinfo {year}
  {2023})}\BibitemShut {NoStop}%
\bibitem [{\citenamefont {Schmidt}\ \emph {et~al.}(2012)\citenamefont
  {Schmidt}, \citenamefont {Erne}, \citenamefont {Nowak}, \citenamefont
  {Sexty},\ and\ \citenamefont {Gasenzer}}]{Schmidt2012}%
  \BibitemOpen
  \bibfield  {author} {\bibinfo {author} {\bibfnamefont {M.}~\bibnamefont
  {Schmidt}}, \bibinfo {author} {\bibfnamefont {S.}~\bibnamefont {Erne}},
  \bibinfo {author} {\bibfnamefont {B.}~\bibnamefont {Nowak}}, \bibinfo
  {author} {\bibfnamefont {D.}~\bibnamefont {Sexty}},\ and\ \bibinfo {author}
  {\bibfnamefont {T.}~\bibnamefont {Gasenzer}},\ }\bibfield  {title} {\bibinfo
  {title} {Non-thermal fixed points and solitons in a one-dimensional {B}ose
  gas},\ }\href {https://doi.org/10.1088/1367-2630/14/7/075005} {\bibfield
  {journal} {\bibinfo  {journal} {New Journal of Physics}\ }\textbf {\bibinfo
  {volume} {14}},\ \bibinfo {pages} {075005} (\bibinfo {year}
  {2012})}\BibitemShut {NoStop}%
\bibitem [{\citenamefont {Lauter}\ \emph {et~al.}(2017)\citenamefont {Lauter},
  \citenamefont {Mitra},\ and\ \citenamefont {Marquardt}}]{Lauter2017}%
  \BibitemOpen
  \bibfield  {author} {\bibinfo {author} {\bibfnamefont {R.}~\bibnamefont
  {Lauter}}, \bibinfo {author} {\bibfnamefont {A.}~\bibnamefont {Mitra}},\ and\
  \bibinfo {author} {\bibfnamefont {F.}~\bibnamefont {Marquardt}},\ }\bibfield
  {title} {\bibinfo {title} {From {K}ardar-{P}arisi-{Z}hang scaling to
  explosive desynchronization in arrays of limit-cycle oscillators},\
  }\bibfield  {journal} {\bibinfo  {journal} {Physical Review E}\ }\textbf
  {\bibinfo {volume} {96}},\ \href {https://doi.org/10.1103/physreve.96.012220}
  {10.1103/physreve.96.012220} (\bibinfo {year} {2017})\BibitemShut {NoStop}%
\bibitem [{\citenamefont {Garc{\'{\i}}a-Morales}\ and\ \citenamefont
  {Krischer}(2012)}]{GarcaMorales2012}%
  \BibitemOpen
  \bibfield  {author} {\bibinfo {author} {\bibfnamefont {V.}~\bibnamefont
  {Garc{\'{\i}}a-Morales}}\ and\ \bibinfo {author} {\bibfnamefont
  {K.}~\bibnamefont {Krischer}},\ }\bibfield  {title} {\bibinfo {title} {The
  complex {G}inzburg{\textendash}{L}andau equation: an introduction},\ }\href
  {https://doi.org/10.1080/00107514.2011.642554} {\bibfield  {journal}
  {\bibinfo  {journal} {Contemporary Physics}\ }\textbf {\bibinfo {volume}
  {53}},\ \bibinfo {pages} {79} (\bibinfo {year} {2012})}\BibitemShut {NoStop}%
\bibitem [{\citenamefont {Diessel}\ \emph {et~al.}(2022)\citenamefont
  {Diessel}, \citenamefont {Diehl},\ and\ \citenamefont
  {Chiocchetta}}]{Diessel2022}%
  \BibitemOpen
  \bibfield  {author} {\bibinfo {author} {\bibfnamefont {O.~K.}\ \bibnamefont
  {Diessel}}, \bibinfo {author} {\bibfnamefont {S.}~\bibnamefont {Diehl}},\
  and\ \bibinfo {author} {\bibfnamefont {A.}~\bibnamefont {Chiocchetta}},\
  }\bibfield  {title} {\bibinfo {title} {Emergent
  {K}ardar{\textendash}{P}arisi{\textendash}{Z}hang phase in quadratically
  driven condensates},\ }\bibfield  {journal} {\bibinfo  {journal} {Physical
  Review Letters}\ }\textbf {\bibinfo {volume} {128}},\ \href
  {https://doi.org/10.1103/physrevlett.128.070401}
  {10.1103/physrevlett.128.070401} (\bibinfo {year} {2022})\BibitemShut
  {NoStop}%
\bibitem [{\citenamefont {Dagvadorj}\ \emph {et~al.}(2021)\citenamefont
  {Dagvadorj}, \citenamefont {Kulczykowski}, \citenamefont {Szyma{\'{n}}ska},\
  and\ \citenamefont {Matuszewski}}]{Dagvadorj2021}%
  \BibitemOpen
  \bibfield  {author} {\bibinfo {author} {\bibfnamefont {G.}~\bibnamefont
  {Dagvadorj}}, \bibinfo {author} {\bibfnamefont {M.}~\bibnamefont
  {Kulczykowski}}, \bibinfo {author} {\bibfnamefont {M.~H.}\ \bibnamefont
  {Szyma{\'{n}}ska}},\ and\ \bibinfo {author} {\bibfnamefont {M.}~\bibnamefont
  {Matuszewski}},\ }\bibfield  {title} {\bibinfo {title} {First-order
  dissipative phase transition in an exciton-polariton condensate},\ }\bibfield
   {journal} {\bibinfo  {journal} {Physical Review B}\ }\textbf {\bibinfo
  {volume} {104}},\ \href {https://doi.org/10.1103/physrevb.104.165301}
  {10.1103/physrevb.104.165301} (\bibinfo {year} {2021})\BibitemShut {NoStop}%
\bibitem [{\citenamefont {Gladilin}\ and\ \citenamefont
  {Wouters}(2017)}]{Gladilin2017}%
  \BibitemOpen
  \bibfield  {author} {\bibinfo {author} {\bibfnamefont {V.~N.}\ \bibnamefont
  {Gladilin}}\ and\ \bibinfo {author} {\bibfnamefont {M.}~\bibnamefont
  {Wouters}},\ }\bibfield  {title} {\bibinfo {title} {Interaction and motion of
  vortices in nonequilibrium quantum fluids},\ }\href
  {https://doi.org/10.1088/1367-2630/aa83a1} {\bibfield  {journal} {\bibinfo
  {journal} {New Journal of Physics}\ }\textbf {\bibinfo {volume} {19}},\
  \bibinfo {pages} {105005} (\bibinfo {year} {2017})}\BibitemShut {NoStop}%
\bibitem [{\citenamefont {Deligiannis}\ \emph {et~al.}(2022)\citenamefont
  {Deligiannis}, \citenamefont {Fontaine}, \citenamefont {Squizzato},
  \citenamefont {Richard}, \citenamefont {Ravets}, \citenamefont {Bloch},
  \citenamefont {Minguzzi},\ and\ \citenamefont {Canet}}]{Deligiannis2022}%
  \BibitemOpen
  \bibfield  {author} {\bibinfo {author} {\bibfnamefont {K.}~\bibnamefont
  {Deligiannis}}, \bibinfo {author} {\bibfnamefont {Q.}~\bibnamefont
  {Fontaine}}, \bibinfo {author} {\bibfnamefont {D.}~\bibnamefont {Squizzato}},
  \bibinfo {author} {\bibfnamefont {M.}~\bibnamefont {Richard}}, \bibinfo
  {author} {\bibfnamefont {S.}~\bibnamefont {Ravets}}, \bibinfo {author}
  {\bibfnamefont {J.}~\bibnamefont {Bloch}}, \bibinfo {author} {\bibfnamefont
  {A.}~\bibnamefont {Minguzzi}},\ and\ \bibinfo {author} {\bibfnamefont
  {L.}~\bibnamefont {Canet}},\ }\bibfield  {title} {\bibinfo {title}
  {{K}ardar{\textendash}{P}arisi{\textendash}{Z}hang universality in discrete
  two-dimensional driven-dissipative exciton polariton condensates},\
  }\bibfield  {journal} {\bibinfo  {journal} {Physical Review Research}\
  }\textbf {\bibinfo {volume} {4}},\ \href
  {https://doi.org/10.1103/physrevresearch.4.043207}
  {10.1103/physrevresearch.4.043207} (\bibinfo {year} {2022})\BibitemShut
  {NoStop}%
\bibitem [{\citenamefont {Comaron}\ \emph {et~al.}(2021)\citenamefont
  {Comaron}, \citenamefont {Carusotto}, \citenamefont {Szyma{\'{n}}ska},\ and\
  \citenamefont {Proukakis}}]{Comaron2021}%
  \BibitemOpen
  \bibfield  {author} {\bibinfo {author} {\bibfnamefont {P.}~\bibnamefont
  {Comaron}}, \bibinfo {author} {\bibfnamefont {I.}~\bibnamefont {Carusotto}},
  \bibinfo {author} {\bibfnamefont {M.~H.}\ \bibnamefont {Szyma{\'{n}}ska}},\
  and\ \bibinfo {author} {\bibfnamefont {N.~P.}\ \bibnamefont {Proukakis}},\
  }\bibfield  {title} {\bibinfo {title} {Non-equilibrium
  {B}erezinskii{\textendash}{K}osterlitz{\textendash}{T}houless transition in
  driven-dissipative condensates},\ }\href
  {https://doi.org/10.1209/0295-5075/133/17002} {\bibfield  {journal} {\bibinfo
   {journal} {Europhysics Letters}\ }\textbf {\bibinfo {volume} {133}},\
  \bibinfo {pages} {17002} (\bibinfo {year} {2021})}\BibitemShut {NoStop}%
\bibitem [{\citenamefont {Ferrier}\ \emph {et~al.}(2022)\citenamefont
  {Ferrier}, \citenamefont {Zamora}, \citenamefont {Dagvadorj},\ and\
  \citenamefont {Szyma{\'{n}}ska}}]{Ferrier2022}%
  \BibitemOpen
  \bibfield  {author} {\bibinfo {author} {\bibfnamefont {A.}~\bibnamefont
  {Ferrier}}, \bibinfo {author} {\bibfnamefont {A.}~\bibnamefont {Zamora}},
  \bibinfo {author} {\bibfnamefont {G.}~\bibnamefont {Dagvadorj}},\ and\
  \bibinfo {author} {\bibfnamefont {M.~H.}\ \bibnamefont {Szyma{\'{n}}ska}},\
  }\bibfield  {title} {\bibinfo {title} {Searching for the
  {K}ardar{\textendash}{P}arisi{\textendash}{Z}hang phase in microcavity
  polaritons},\ }\bibfield  {journal} {\bibinfo  {journal} {Physical Review B}\
  }\textbf {\bibinfo {volume} {105}},\ \href
  {https://doi.org/10.1103/physrevb.105.205301} {10.1103/physrevb.105.205301}
  (\bibinfo {year} {2022})\BibitemShut {NoStop}%
\bibitem [{\citenamefont {Dagvadorj}\ \emph {et~al.}(2015)\citenamefont
  {Dagvadorj}, \citenamefont {Fellows}, \citenamefont {Matyja{\'{s}}kiewicz},
  \citenamefont {Marchetti}, \citenamefont {Carusotto},\ and\ \citenamefont
  {Szyma{\'{n}}ska}}]{Dagvadorj2015}%
  \BibitemOpen
  \bibfield  {author} {\bibinfo {author} {\bibfnamefont {G.}~\bibnamefont
  {Dagvadorj}}, \bibinfo {author} {\bibfnamefont {J.}~\bibnamefont {Fellows}},
  \bibinfo {author} {\bibfnamefont {S.}~\bibnamefont {Matyja{\'{s}}kiewicz}},
  \bibinfo {author} {\bibfnamefont {F.}~\bibnamefont {Marchetti}}, \bibinfo
  {author} {\bibfnamefont {I.}~\bibnamefont {Carusotto}},\ and\ \bibinfo
  {author} {\bibfnamefont {M.}~\bibnamefont {Szyma{\'{n}}ska}},\ }\bibfield
  {title} {\bibinfo {title} {Nonequilibrium phase transition in a
  two-dimensional driven open quantum system},\ }\bibfield  {journal} {\bibinfo
   {journal} {Physical Review X}\ }\textbf {\bibinfo {volume} {5}},\ \href
  {https://doi.org/10.1103/physrevx.5.041028} {10.1103/physrevx.5.041028}
  (\bibinfo {year} {2015})\BibitemShut {NoStop}%
\end{thebibliography}

\providecommand{\noopsort}[1]{}\providecommand{\singleletter}[1]{#1}%

\end{document}